\documentclass[12pt]{article}
																			 																																																												\pdfoutput=1	
\usepackage{amsmath,amssymb}
\usepackage{graphicx}
\usepackage[pdftex]{hyperref}
\usepackage{epstopdf}
\usepackage{xcolor}
\DeclareGraphicsExtensions{.eps,.bmp,.wmf,.jpeg,.pdf}
\numberwithin{equation}{section}
\def\be{\begin{equation}}
\def\ee{\end{equation}}

\def\bea{\begin{eqnarray}}
\def\eea{\end{eqnarray}}
\title{Modified gravity with disappearing cosmological constant}
\author{L. N. Granda\thanks{luis.granda@correounivalle.edu.co}\\ {\small\it Departamento de Fisica, Universidad del Valle}\\{\small\it A.A. 25360, Cali, Colombia}}
\date{}
\begin{document}
\bibliographystyle{alpha}
\maketitle


\begin{abstract}
\noindent 
New corrections to General Relativity are considered in the context of modified $f(R)$ gravity, that satisfy cosmological and local gravity constraints. The proposed models behave asymptotically as $R-2\Lambda$ at large curvature and show the vanishing of the cosmological constant at the flat spacetime limit.
The chameleon mechanism and thin shell restrictions for local systems were analyzed, and bounds on the models were found. The steepness of the deviation parameter $m$ at late times leads to measurable signal of scalar-tensor regime in matter perturbations, that allows to detect departures form the $\Lambda$CDM model. The theoretical results for the evolution of the weighted growth rate $f\sigma_8(z)$, from the proposed models, were analyzed.
\end{abstract}

\section{Introduction}

\noindent Among all models of dark energy, the $\Lambda$CDM is the simplest and the most accurate in terms of consistency with observational data (for review see \cite{copeland, sahnii, padmanabhan, sergeiod}). However its non-dynamic behavior gives rise to a single possible cosmological scenario in which the fine-tuning problem cannot be solved. This motivates the development of different approaches to the dark energy problem with models that have dynamical nature and avoid the introduction of a cosmological constant. Among these models, modified gravity $f(R)$ models stand out, especially after the recent discovery of gravitational waves \cite{ligo} and the measurement of their speed with great accuracy, which led to the discarding of several scalar-tensor models and models belonging to the class of Horndeski or Galilean theories \cite{horndeski, nicolis, deffayet}. An attractive feature 
of modified gravity models is that they lead to cosmic accelerated expansion without introducing a dark energy matter component. These $f(R)$ models contain non-linear in $R$ corrections to the General Relativity that must pass various restrictions ranging from cosmological to those imposed by local gravity phenomena (see \cite{sodintsov1, sotiriou, tsujikawa01, tsujikawa0, odinnojiri, odinoiko4} for reviews). Most $f(R)$ models pass cosmological constraints, but the main obstacle for being successful is the fulfillment with the more stringent local gravity constraints. Compliance with these local constraints renders many models indistinguishable from $\Lambda$CDM and probably it is not possible to distinguish them from $\Lambda$CDM through observations, at least with the current precision available. 
So it is important to consider models that maintain a balance between behaving like General Relativity (GR) in local phenomena and showing signals of modified gravity at other scales, which can be measured with the next improvement in observational capacity. Thus for instance, there can be differences in the dynamics of perturbations that lead to interesting signatures that can be observed in the near future. \\
Citing just a few among the large amount of work done in $f(R)$ gravity \cite{capozziello, capozziello1, sodintsov, nojiri5, carroll, nojiriodin, nojiri, elizalde, troisi, allemandi, koivisto, msami, barrow1a, faraoni, brevik, koivisto1, sotiriou1, nojiri1, dobado, sodintsov2, anthoni, nojiri2, faraoni1, song, bean, olmo, amendola1, barrow1, fay, faraoni2, hu, nojiri6, tsujikawa1, nojiri7, sergeid2, bamba1, elizalde1, sergeioiko, sergeisaez, sergeisaez1}, it can be highlighted that the most studied corrections to the Einstein gravity are those containing positive and negative powers of curvature, for which it was found that positive powers of curvature are important at early times and can lead to successful inflation like the Starobinsky $R^2$  model \cite{starobinsky}, while negative powers of curvature can give consistent late time cosmological behavior but contain instabilities that prohibit the formation of a matter dominated era, and are also inconsistent with solar system tests  \cite{dolgov, lucamendola, nojiri5, olmo1, sodintsov2, anthoni, faraoni1}.
Cosmological constraints on $f(R)$ models using different observational data were performed in \cite{troisi1, mota, dev, schmidt, lombriser, nesseris, nunes}. Solar system constraints and chameleon mechanism have been studied in \cite{olmo, olmo1,hu, chiba, lamendola5, tsujikawa, brax}, and $f(R)$ models that can satisfy both cosmological and local gravity constraints have been proposed in \cite{hu, astarobinsky, appleby1, sergeid2, elizalde1, granda1, granda3}. \\
In the present paper we propose $f(R)$ models that satisfy the stability conditions $f'(R)>0,\; f''(R)>0$, comply with cosmological and local gravity constraints and can lead to signals of scalar-tensor regime measurable at late times. 
Corrections to the GR  of two types are considered. Corrections of the form $e^{-g_1(R)}$ where the function $g_1(R)$ is a positive definite function that satisfies the asymptotic behavior, $g_1(R\to \infty)\to 0$ and $g_1(R\to 0)\to \infty$.
The second type of corrections are of the form $(1-e^{-g_2(R)})$, where $g_2(R)$ is a positive definite function that satisfies  $g_2(R\to \infty)\to \infty$ and $g_2(R\to 0)\to 0$.  The first limit leads to an effective cosmological constant while the second leads to disappearing cosmological constant in the flat space time limit.
Hence the accelerated expansion is explained as a geometrical effect.\\ 
This paper is organized as follows. In section 2 we present the general features of the $f(R)$ models. In section 3 we present some models that give viable cosmologies, and analyze their behavior under  large curvature regime and at late times. In section 4 we analyze the restrictions from matter density perturbations. Some discussion is given un section 5.

\section{General field equations and constraints}
The modified gravity is described by a general action of the form
\be\label{eq1}
S=\int d^4x\sqrt{-g}\left[\frac{1}{2\kappa^2}f(R)+{\cal L}_m\right]
\ee
where $\kappa^2=8\pi G$, $f(R)$ is a function of curvature that contains the linear Einstein term and non-linear corrections to it, and ${\cal L}_m$ is the Lagrangian density for the matter component which satisfies the usual conservation equation. For the flat Friedmann-Robertson-Walker metric the equations of motion are given by
\be\label{eq2a}
3H^2F=\frac{1}{2}\left(RF-f\right)-3H\dot{F}+\kappa^2\rho
\ee
and
\be\label{eq2b}
-2\dot{H}F=\ddot{F}-H\dot{F}+\kappa^2\left(\rho+p\right)
\ee
where dot represents derivative with respect to cosmic time, $F=f_{,R}=\partial f/\partial R$ and $\rho$ and $p$ are the energy density and pressure for the matter component represented as a perfect fluid  (in what follows we will use indistinctly $f_{,R}$ or $F=f_{,R}$). The field equation (\ref{eq2a}) can be written in more compact form by defining the effective energy density as follows
\be\label{effdensity}
H^2=\frac{\kappa^2}{3}\rho_{eff},
\ee
where
\be\label{effdensity1}
\rho_{eff}=\frac{1}{F}\left[\frac{1}{2\kappa^2}\left(RF-f-6H\dot{F}\right)+\rho\right]
\ee
The Eqs. (\ref{eq2a}) and (\ref{eq2b}) lead to the following effective equation of state (EoS)
\be\label{eos}
w_{eff}=-1-\frac{2\dot{H}}{3H^2}=-1+\frac{\ddot{F}-H\dot{F}+\kappa^2\left(\rho+p\right)}{\frac{1}{2}\left(RF-f\right)-3H\dot{F}+\kappa^2\rho},
\ee
where $\rho$ and $p$ include both matter and radiation components, i.e. $\rho=\rho_m+\rho_r$ and $p=p_m+p_r$. Defining the modified density parameters $\Omega_m$ and $\Omega_r$ as
\be\label{densities}
\Omega_m=\frac{\kappa^2\rho_m}{3FH^2},\;\;\; \Omega_r=\frac{\kappa^2\rho_r}{3FH^2},
\ee
one can write the DE equation of state as follows \cite{amendola1}
\be\label{deeos}
w_{DE}=\frac{w_{eff}-F/(3F_0)\Omega_r}{1-F/F_0\left(\Omega_m+\Omega_r\right)}
\ee
where $F_0$ is the current value of $F$ that is used to rewrite Eqs. (\ref{eq2a}) and (\ref{eq2b}) as
\be\label{deeos1}
3F_0H^2=\kappa^2\left(\rho_m+\rho_r+\rho_{DE}\right)
\ee
and
\be\label{deeos2}
-2F_0\dot{H}=\kappa^2\left(\rho_m+\frac{4}{3}\rho_r+\rho_{DE}+p_{DE}\right).
\ee
$w_{DE}$ can also be written in terms of the redshift as
\be\label{deeos3}
w_{DE}=\frac{1}{3}\frac{(1+z)\frac{d\tilde{H}^2}{dz}-3\tilde{H}^2-\Omega_{r0}(1+z)^4}{\tilde{H}^2-\Omega_{m0}(1+z)^3-\Omega_{r0}(1+z)^4},
\ee
where $\tilde{H}=H/H_0$ and the subscript "0" stands for present values.\\
In general, the function $f(R)$ can be written as the linear term that describes the Einstein gravity plus a non-linear function of $R$ that describe the deviations from Einstein gravity that must be negligible (compared to the curvature) in the early universe and become relevant at late times to account for accelerated cosmic expansion. \\
Any suitable $f(R)$ model must comply with the absence of ghost instabilities  and must be stable under matter perturbations at high curvature regime \cite{dolgov, faraoni1}, that are satisfied if the conditions
\be\label{stability}
\frac{\partial f(R)}{\partial R}>0, \;\;\; \frac{\partial^2 f(R)}{\partial R^2}>0
\ee
take place throughout the whole period of evolution of the universe. In modified gravity, due to the non-linear correction to $R$, there is a propagating scalar degree of freedom $f_{,R}$ whose dynamics follows from the trace equation given by
\be\label{trace}
\Box f_{,R}(R)=\frac{1}{3}\left(2f(R)-Rf_{,R}(R)\right)+\frac{\kappa^2}{3}\left(3p-\rho\right)=\frac{dV_{eff}}{df_{,R}}.
\ee
where the right hand side of this equation is represented as the derivative of an effective potential $V_{eff}$ with respect to the scalar field $f_{,R}$. Then the mass of 
$f_{,R}$ can be defined as
\be\label{mass}
M^2=\frac{d^2 V_{eff}}{f_{,R}^2}=\frac{1}{3}\left(\frac{f_{,R}}{f_{,RR}}-R\right).
\ee
Since viable models satisfy $f_{,R}\approx 1$, then and at high curvature (in matter epoch for instance) $Rf_{,RR}<<1$, and this mass can be approximated as
\be\label{mass1}
M^2\simeq \frac{1}{3f_{,RR}}
\ee
This mass allows to define the corresponding Compton wavelength, $\lambda_C=2\pi/M$, that mediates the interaction due to the extra scalar degree of freedom also called scalaron. In regions of high density (compared to background density) where GR is dominant, the scalaron mass acquires large values (compared to the corresponding background value) giving rise to the so called chameleon mechanism \cite{amanda1,amanda2} which will be discussed in the next section.\\
On the other hand, the cosmological viability of an $f(R)$ model imply the consistency with all observational evidence on late time accelerated expansion and also consistency with the high redshift universe where the GR is valid. For its analysis it is useful to resort to the parameters $r$ and $m$ defined as
\be\label{r-m}
r=-\frac{Rf_{,R}}{f},\;\;\; m=\frac{Rf_{,RR}}{f_{,R}},
\ee
that characterize de deviation from the $\Lambda$CDM model. In fact the matter dominant era corresponds to $r=-1$ and $m=0$, where the GR is dominant and    
$ \Lambda$CDM corresponds to the line $m=0$. Then the consistency with observations at high redshift imply that $m<<1$. Using $m$ we can write $M^2$ given in (\ref{mass}) as
\be\label{mass2}
M^2=\frac{R}{3m}\left(1-m\right).
\ee
then the approximation (\ref{mass1}), which from (\ref{r-m}) also imply the condition $m<<1$, can be expressed as
\be\label{mass3}
M^2\simeq \frac{R}{3m}.
\ee
The most stringent constraints are related to the local gravity systems where the curvature is much larger than that of the background. In local systems, as well as at high curvature, the model must be practically indistinguishable from GR, which implies for an $f(R)$ model that $f_{,R}(R_{\ell})\simeq 1$ (or $\lim_{R\to \infty}f(R)/R=1$) and $f_{,RR}(R_{\ell})<<R_{\ell}^{-1}$, where $R_{\ell}$ is the typical curvature of the local system which satisfies $R_{\ell}>>R_b$, where $R_b$ is the background curvature. This also applies when $R>>R_0$ ($R_0$ is the current curvature) and restrictions from Big Bang nucleosynthesis and the Cosmic Microwave Background appear. Note that the cosmological value of the product $Rf_{,RR}$ at current, low curvature Universe, is not necessarily too close to zero, since the viability of $f(R)$ models allows current values  of the deviation parameter $m(R_0) \lesssim {\cal {O}}(1)$.\\

\noindent {\bf Background evolution}\\
To solve numerically the field  equations we use the variables introduced in \cite{hu,bamba1}
\be\label{back1}
y_{H}=\frac{H^2}{\mu^2}-a^{-3},\;\; y_R=\frac{R}{\mu^2}-3a^{-3}
\ee
and work with the $e$-fold variable $\ln a$, where (') indicates $d/d\ln a$. Note that if we assume 
\be\label{back2}
\mu^2=\frac{1}{3}\kappa^2\rho_{m0},
\ee
then we can write $y_H=\rho_{DE}/\rho_{m0}$, from which it becomes clear that at high redshift, where matter dominates, $y'_H\sim 0$. Taking into account $$R=6\left(2H^2+HH'\right)$$ and (\ref{eq2a}) we find the following equation for $y_H$ (after decoupling from $y_R$)
\be\label{back3}
y''_H+J_1y'_H+J_2 y_H+J3=0
\ee
where
\be\label{back4}
J_1=4+\frac{1-f_{,R}}{6\mu^2(y_H+a^{-3})f_{,RR}}
\ee
\be\label{back5}
J_2=\frac{2-f_{,R}}{3\mu^2(y_H+a^{-3})f_{,RR}}
\ee
\be\label{back6}
J_3=-3a^{-3}-\frac{1}{6\mu^2(y_H+a^{-3})f_{,RR}}\left[(1-f_{,R})a^{-3}+\frac{1}{3}\frac{R-f}{\mu^2}\right]
\ee
and for $y_{R}$ it is found
\be\label{back7}
y_R=3\left(y'_{H}+4y_{H}\right).
\ee
The equations (\ref{eq2a}) and (\ref{eq2b}) can be written in the standard form 
\be\label{back8}
3H^2=\kappa^2\left(\rho_m+\rho_{DE}\right),\;\; 2\dot{H}=-\kappa^2\left(\rho_m+\rho_{DE}+p_{DE}\right),
\ee
where 
\be\label{back9}
\rho_{DE}=\frac{3}{\kappa^2}H^2-\rho_m=\frac{3}{\kappa^2}\left[\frac{1}{6}\left(Rf_{,R}-f\right)-H^2\left(f_{,R}+R' f_{,RR}-1\right)\right]
\ee
which allows to write the EoS of dark energy and the effective EoS in terms of $y_H$ and $y_R$ as
\be\label{back10}
w_{DE}=-1-\frac{1}{3}\frac{y'_H}{y_H},\;\;\;  w_{eff}=-\frac{3y_H+y'_H}{3\left(y_H+a^{-3}\right)}
\ee
The background evolution can be analyzed by solving the Eq. (\ref{back3}) numerically, which allows to find $y_H$ as function of the redshift.   

\section{Viable $f(R)$ models.}
Here we discuss some $f(R)$ models that meet all required conditions of stability, cosmological viability, satisfy local gravity constraints and leave their trace on the evolution of matter density perturbations. We introduce the following models
\subsection*{Corrections of the type $e^{-g_1(R)}$}
\noindent We can define a class of modified gravity models that are represented by functions of the form
\be\label{general-f}
f(R)=R-\lambda\mu^2 e^{-g_1(R)}
\ee
where the function $g(R)$ is positive definite and satisfies the asymptotic behavior 
\be\label{general-f1}
\lim_{R\to \infty}g_1(R) =0,\;\;\; \lim_{R\to 0}g_1(R)=\infty.
\ee
These type of models lead to the absence of cosmological constant in the flat space-time limit. 
The simplest choice for $g(R)$ that satisfies these conditions is the monomial
\be\label{general-f2}
g_1(R)=\left(\frac{\mu^2}{R}\right)^{\eta},
\ee
which corresponds to the model proposed in \cite{granda1,granda3}.
The next simple case is given by the following function
\be\label{model3}
g_1(R)=\alpha\ln\left[1+\left(\frac{\mu^2}{R}\right)^{\eta}\right]
\ee
with  $\eta>0$ and $\alpha>0$, which leads the $f(R)$ model
\be\label{model2}
f(R)=R-\lambda \mu^2\left[1+\left(\frac{\mu^2}{R}\right)^{\eta}\right]^{-\alpha}
\ee
where $\lambda>0$.
This model behaves asymptotically as
\be\label{limmodel2}
\lim_{R\to \infty}\left(f(R)-R\right) =-\lambda\mu^2,\;\;\; \lim_{R\to 0}f(R)=0.
\ee
So, the cosmological constant disappears in the flat spacetime limit. In the regime $\mu^2<<R$ this model behaves as HS and Starobinsky models 
\be\label{model2a}
f(R)\simeq R-\lambda\mu^2\left(1-\alpha\left(\frac{\mu^2}{R}\right)^{\eta}\right)
\ee
and also coincides with the three-parameter HS model ($c_2=1$ in HS \cite{hu}) for $\alpha=1$. 
The model (\ref{model2})  can also be written in the form
\be\label{general-f4}
f(R)=R-\lambda\mu^2\frac{\left(\frac{R}{\mu^2}\right)^{\alpha\eta}}{\left[\left(\frac{R}{\mu^2}\right)^{\eta}+1\right]^{\alpha}},
\ee
To analyze the stability conditions we write the first and second derivatives of (\ref{model2}) 
\be\label{1model2}
f_{,R}=1-\alpha\lambda\eta\left(\frac{\mu^2}{R}\right)^{\eta+1}\left[1+\left(\frac{\mu^2}{R}\right)^{\eta}\right]^{-\alpha-1}
\ee
\be\label{2model2}
f_{,RR}=\frac{\alpha\lambda\eta}{\mu^2}\left(\frac{\mu^2}{R}\right)^{\eta+2}\left[1+\left(\frac{\mu^2}{R}\right)^{\eta}\right]^{-\alpha-2}\left(1+\eta+(1-\alpha\eta)\left(\frac{\mu^2}{R}\right)^{\eta}\right).
\ee
 with $\eta>0$ and $\alpha>0$, a sufficient condition for $f_{,R}>0$ is the following
\be\label{stabil1}
\alpha\lambda\eta<\left(\frac{R}{\mu^2}\right)^{\eta+1}
\ee
and the condition $f_{,RR}>0$ leads to
\be\label{stabilitya}
1+\eta+(1-\alpha\eta)\left(\frac{\mu^2}{R}\right)^{\eta}>0.
\ee
This inequality is satisfied, independently of $R$, in the cases $\alpha=1/\eta$ or $\alpha\eta<1$. Depending on $R$, $f_{,RR}>0$ is satisfied if
\be\label{stabilityb}
\eta\alpha>1\;\;\; and \;\;\; \frac{\mu^2}{R}<\left(\frac{\eta+1}{\alpha\eta-1}\right)^{1/\eta}.
\ee
The de Sitter curvature from $r(R_{ds})=-2$ can be found by fixing  $\lambda$, which gives  ($R_{ds}=\mu^2 y_{ds}$)
\be\label{lambda2-ds}
\lambda=\frac{y_{ds}\left(1+y_{ds}^{-\eta}\right)^{\alpha+1}}{2+(2-\alpha\eta)y_{ds}^{-\eta}}.
\ee
The condition for $\lambda>0$ is accomplished if 
\be\label{alpha}
0<\alpha\le\frac{2}{\eta}
\ee
which is valid for any $y_{ds}$, or depending on $y_{ds}$
\be\label{alpha1}
\alpha>\frac{2}{\eta},\;\;\; \&\;\;\; y_{ds}>\left(\frac{\alpha\eta-2}{2}\right)^{1/\eta}
\ee
Given $\eta>1$ and assuming that  $y_{ds}>>1$ (as in fact takes place for the initial conditions we will use), $\lambda$ can be approximated as
\be\label{lambda2a-ds}
\lambda\approx \frac{1}{2}y_{ds}.
\ee
Replacing (\ref{lambda2-ds})  in (\ref{model2}) and using the Eqs. (\ref{r-m}) we find  (setting $R=\mu^2 y$)
\be\label{m2y}
m=\frac{\alpha\eta y_{ds}\left(1+y_{ds}^{-\eta}\right)^{\alpha+1}y^{-\eta}\left((\alpha\eta-1)y^{-\eta}-\eta-1\right)}{\left(1+y^{-\eta}\right)\left[\alpha\eta y_{ds}\left(1+y_{ds}^{-\eta}\right)^{\alpha+1}y^{-\eta}+\left((\alpha\eta-2)y_{ds}^{-\eta}-2\right)y\left(1+y^{-\eta}\right)^{\alpha+1}\right]},
\ee
\be\label{r2y}
r=-\frac{y\left[\alpha\eta y_{ds}\left(1+y_{ds}^{-\eta}\right)^{\alpha+1}y^{-\eta-1}\left(1+y^{-\eta}\right)^{-\alpha-1}+\left(\alpha\eta-2\right)y_{ds}^{-\eta}-2\right]}{y_{ds}\left(1+y_{ds}^{-\eta}\right)^{\alpha+1}\left(1+y^{-\eta}\right)^{-\alpha}+\left(\left(\alpha\eta-2\right)y_{ds}^{-\eta}-2\right)y}.
\ee
To find the stability condition at the de Sitter point, we evaluate $m(y_{ds})$ obtaining
\be\label{mds}
m(y_{ds})=\frac{\alpha\eta y_{ds}^{-\eta}\left(1+\eta+(1-\alpha\eta) y_{ds}^{-\eta}\right)}{2\left(1+ y_{ds}^{-\eta}\right)\left(1+(1-\alpha\eta) y_{ds}^{-\eta}\right)}.
\ee
Then, the condition of stability ($0<m(r=-2)\le 1$) can be accomplished, consistently with (\ref{alpha1}), if the following inequalities are satisfied
\be\nonumber
\eta>1\;\; \&\;\; \alpha>\frac{2}{\eta}\;\; \&\;\; 
\ee
\be\nonumber
y_{ds}\ge \left[\frac{2\left(\alpha^2\eta^2-3\alpha\eta+2\right)}{\alpha\eta^2+3\alpha\eta-4-\sqrt{\alpha^2\eta^4+6\alpha^2\eta^3+\alpha^2\eta^2-8\alpha\eta^2}}\right]^{1/\eta}.
\ee
\noindent Numerical analysis shows that models (\ref{model2}) with  $\eta<1$ satisfy cosmological and local gravity constraints, but these last constraints imply too small values of $m(r)$ at current or late times ($m<<10^{-6}$), making it very difficult to detect measurable differences with the $\Lambda$CDM model. More attractive are the results obtained in the case $\eta>1$.\\
From (\ref{model2}), (\ref{model2a}) it follows that $\lambda\mu^2$ should be compared to the observed value of the cosmological constant 
\be\label{current-l2}
\lambda\mu^2\approx 2\Lambda.
\ee
On the other hand, using the density parameter for the cosmological constant, $\Omega_{\Lambda}=\Lambda/(3H_0^2)$, we arrive at the following relation from (\ref{lambda2a-ds})
\be\label{current-l2b}
\lambda\mu^2\approx \frac{1}{2}y_{ds}\mu^2=\frac{1}{2}R_{ds}\approx 6H_0^2\Omega_{\Lambda}\approx \frac{1}{2}\Omega_{\Lambda}R_0.
\ee
Taking the value (\ref{back2}) for the scale $\mu^2$, in Fig. 1 we show the background of the model (\ref{model2}) evolution for some cases. 

\begin{figure}[h]
\begin{center}
\includegraphics[scale=0.59]{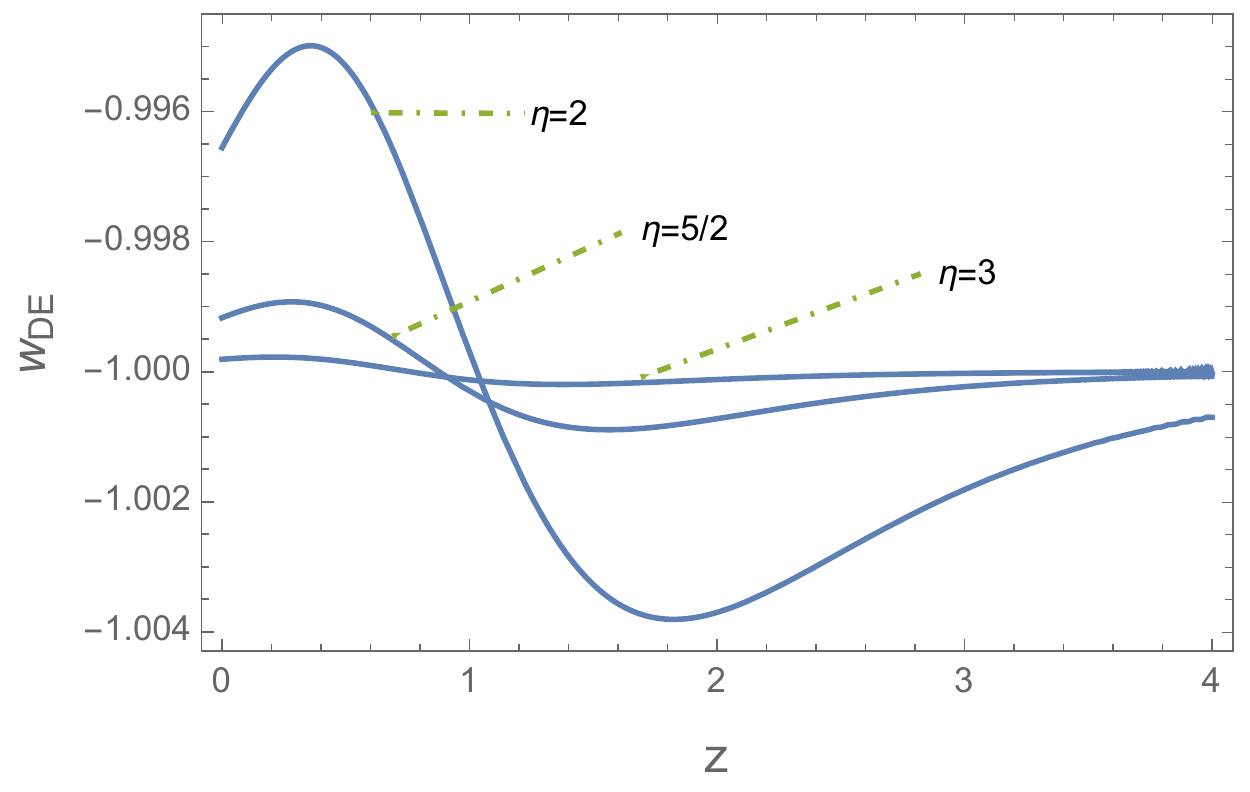}
\includegraphics[scale=0.59]{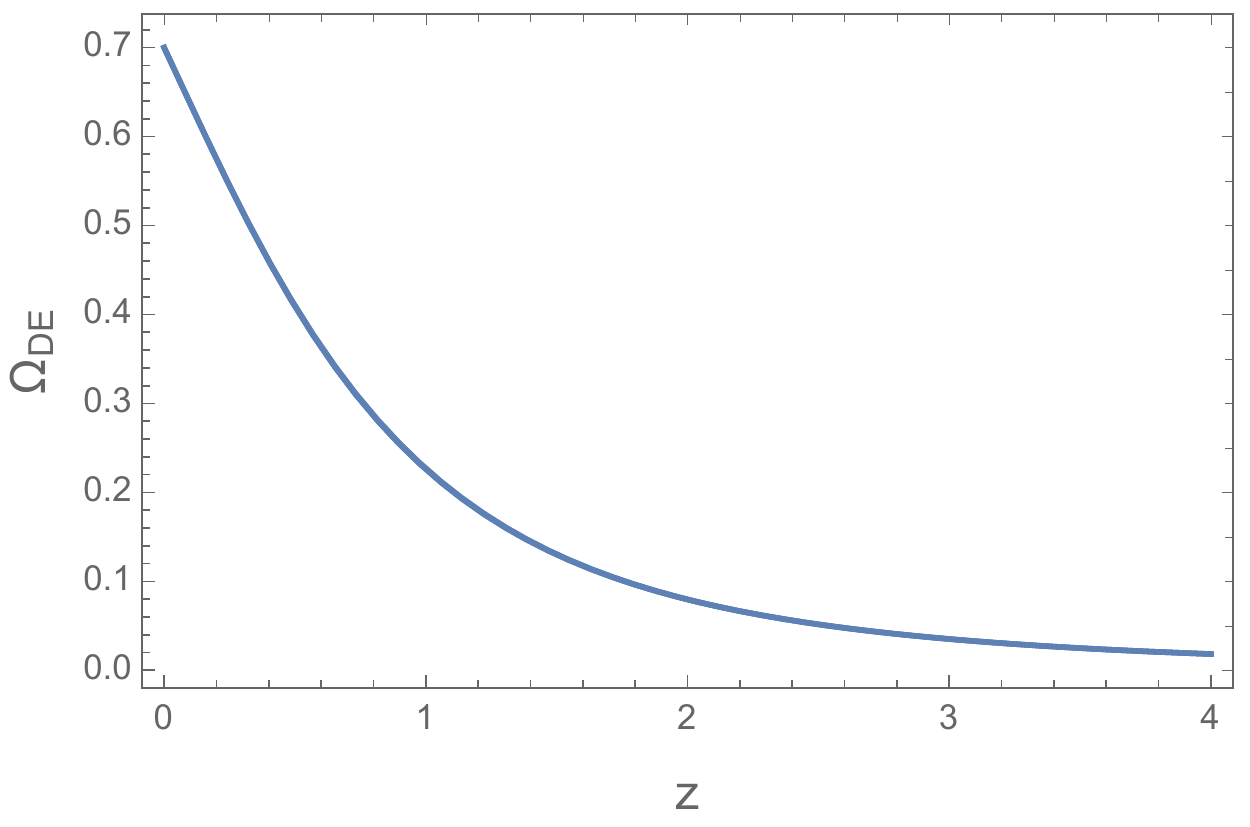}
\caption{The evolution of $w_{DE}$ and $\Omega_{DE}$ for the (\ref{model2}), assuming $\Omega_{m0}=0.3$. For all cases $\alpha=2$. The initial redshift is $z_i=10$ for $\eta=2$, $z_i=8$ for $\eta=5/2$ and $z_i=6.39$ for $\eta=3$. In all cases the evolution of $\Omega_{DE}$ is indistinguishable from that of $\Lambda$CDM.}
\label{fig1}
\end{center}
\end{figure}
\noindent Comparing the model (\ref{model2}) with the HS \cite{hu} model, we can see that both models depend on four parameters. The difference lies in the fact that in the model (\ref{model2}) the parameter $\alpha$ acts as a power, while in HS the parameter $c_2$ is a coefficient. Only making  $\alpha=c_2=1$ the two models coincide. But even if $\alpha\ne c_2$, with the appropriate choice of $\alpha$ and $c_2$, both models give very similar results provided $\mu^2<<R$. \\
During matter dominated epoch or at high-curvature regime, when $R>>\mu^2$, a good approximation for the deviation parameter $m$ will be given by  the expression  
\be\label{maproxmodel2}
m\approx \frac{\alpha\eta(\eta+1)y_{ds}}{2y^{\eta+1}},
\ee
and from (\ref{r2y}), $r$ simplifies to 
\be\label{rapprox1}
r\approx - \frac{2y}{2y-y_{ds}},
\ee
which allows to write explicitly $m(r)$ as 
\be\label{mapprox1}
m(r)\approx \frac{1}{2}\alpha\eta(\eta+1)y_{ds}\left(\frac{2(r+1)}{y_{ds}r}\right)^{\eta+1}
\ee
To have an estimation of the effect $\alpha$ in the behavior $m$ at late times, we consider the current value of the background curvature $R_0$ which is given by (using (\ref{back2}))
\be\label{R0}
R_0\approx \frac{3\mu^2}{\Omega_{m0}}\left(4-3\Omega_{m0}\right).
\ee
Then, from (\ref{maproxmodel2}) or (\ref{mapprox1}) it follows that (using (\ref{current-l2b}))
\be\label{mds1}
m(y_{0})\sim 6\alpha\eta(\eta+1)\frac{1-\Omega_{m0}}{\Omega_{m0}}\left(\frac{12}{\Omega_{m0}}-9\right)^{-(\eta+1)}.
\ee
So that at late times $\alpha$ may affect the order of magnitude of $m$ impacting in the steepness of $m(r)$. On the other hand, and given that $y>>y_{ds}$, in a high curvature regime the parameter $\alpha$ is not so relevant in the expression (\ref{maproxmodel2}) (unless it is large enough), so the power $\eta$ becomes the dominant parameter.\\
Another case of the model (\ref{model2}) is obtained by setting $\alpha=1/\eta$, giving 
\be\label{model3b}
f(R)=R-\frac{\lambda R}{\left[\left(\frac{R}{\mu^2}\right)^{\eta}+1\right]^{1/\eta}}.
\ee
The parameters $m$ and $r$ are given by the simple expressions
\be\label{mmodel3b}
m=\frac{\lambda(\eta+1)y^{\eta}}{(y^{\eta}+1)\left(y^{\eta}+1\right)^{\frac{\eta+1}{\eta}}-\lambda},\;\; r=-\frac{1-\lambda\left(y^{\eta}+1\right)^{-\frac{\eta+1}{\eta}}}{1-\lambda\left(y^{\eta}+1\right)^{-\frac{1}{\eta}}},
\ee
where $\lambda$ is fixed by the de Sitter solution
\be\label{lmodel3b}
\lambda=\frac{\left(y_{ds}^{\eta}+1\right)^{\frac{\eta+1}{\eta}}}{2y_{ds}^{\eta}+1}
\ee
Note that $\lambda>0$ without restrictions. From the previous analysis, applied to the case $\alpha=1/\eta$, it can be seen that a sufficient condition for $f_{,R}>0$ is accomplished if $\lambda<\left(R/\mu^2\right)^{\eta+1}$  ($R>\mu^2$ throughout cosmological evolution), and the condition $f_{RR}>0$ always takes place without restrictions, for any $R$. The stability condition at the de Sitter point ($0<m(r=-2)\le 1$) is satisfied for $\eta>1$ and $y_{ds}\ge \left(\frac{\eta-1}{2}\right)^{1/\eta}$.\\ 
If we limit ourselves to the restrictions imposed by local gravity and the bound $m(z\approx 0)\gtrsim 3\times 10^{-6}$ (see matter perturbations below), then the only interesting value left for $\eta$ is $\eta=3$. Lower values will not satisfy local gravity restrictions, and higher values lead to $m(z\approx 0)< 10^{-6}$, i.e. below the proposed bound.\\
It is worth noticing that the HS model \cite{hu} corresponds to the function
\be\label{model4}
g_1(R)=\ln\left[1+\lambda_1\left(\frac{\mu^2}{R}\right)^{\eta}\right],
\ee

 \subsection*{Corrections of the type $(1-e^{-g_2(R)})$}
Another important class of models can be generated by functions $fR)$ of the type
\be\label{moldel-l}
 f(R)=R-\lambda\mu^2\left(1-e^{-g_2(R)}\right)
 \ee
where the function $g_2(R)$ satisfies the asymptotic limits
\be\label{limitmodel-l}
\lim_{R\to \infty}g_2(R) =\infty,\;\;\; \lim_{R\to 0}g_2(R)=0.
\ee
{\bf Model 1}.\\
The simplest choice for these models is the function
\be\label{model-l0} 
 g_2(R)=\left(\frac{R}{\mu^2}\right)^{\eta},
 \ee
which  leads to
\be\label{model-l1}
 f(R)=R-\lambda\mu^2\left[1-e^{-\left(\frac{R}{\mu^2}\right)^{\eta}}\right],
 \ee
where $\lambda>0$, $0<\eta<1$ and $\mu^2<R$. The best known example is the exponential model \cite{sergeid2, linder1} that corresponds to $\eta=1$. As in the case of models (\ref{general-f}), these models lead to the disappearance of cosmological constant in the flat space-time limit. 
As will be shown below, the rapid zero trend of the exponential model ($\eta=1$) can be substantially attenuated considering models with $\eta<1$, while all local gravity and cosmological restrictions are respected.\\
The first and second derivatives are
\be\label{1model-l}
f_{,R}=1-\eta\lambda\left(\frac{R}{\mu^2}\right)^{\eta-1}e^{-\left(\frac{R}{\mu^2}\right)^{\eta}}
\ee
\be\label{2model-l}
f_{,RR}=\frac{\eta\lambda}{\mu^2}\left(1-\eta+\eta\left(\frac{R}{\mu^2}\right)^{\eta}\right)\left(\frac{R}{\mu^2}\right)^{\eta-2}e^{-\left(\frac{R}{\mu^2}\right)^{\eta}}
\ee
The stability condition $f_{,RR}>0$ is automatically satisfied for $\eta<1$. A sufficient condition for $f_{,R}>0$, given that $\mu^2<R$ and $\eta<1$, is satisfied if $\eta\lambda<(R/\mu^2)^{1-\eta}$.
For $r$ and $m$ we find, setting $R=y\mu^2$
\be\label{rmodel-l}
r=\frac{\eta\lambda\lambda_1 y^{\eta}-y e^{-y^{\eta}}}{\left(y-\lambda\right)e^{y^{\eta}}+\lambda\mu^2},
\ee
\be\label{mmodel-l}
m=\frac{\eta\lambda\left(1-\eta+\eta y^{\eta}\right)y^{\eta}}{ye^{ y^{\eta}}-\eta\lambda y^{\eta}}
\ee
Note that $m>0$ provided that the conditions for $f_{,R}>0$ hold. It also follows for $\eta,\lambda>0$ that 
\be\label{matter}
\lim_{y\to\infty}r=-1,\;\;\; \lim_{y\to\infty} m=0
\ee
showing that all trajectories contain the matter-dominated point $(r=-1,m=0)$.
The de Sitter attractor is fixed by solving the equation $r(y_{ds})=-2$ with respect to $\lambda$, which gives
 \be\label{dsmodel-l}
 \lambda=\frac{y_{ds} e^{y_{ds}^{\eta}}}{2e^{ y_{ds}^{\eta}}-\eta y_{ds}^{\eta}-2}.
\ee
From this expressions follows that $\lambda>0$ given that $0<\eta<1$. Replacing $\lambda$ in (\ref{mmodel-l}) and evaluating at  $y_{ds}$ gives
\be\label{mds1model-l}
m(y_{ds})=\frac{\eta y_{ds}^{\eta}\left(1-\eta+\eta y_{ds}^{\eta}\right)}{2e^{y_{ds}^{\eta}}-2\eta y_{ds}^{\eta}-2}.
\ee
To check the conditions of stability at de Sitter point, $0<m(r=-2)\le 1$, it is useful to expand the exponential, that yields
\be\label{mds2model-l}
m(y_{ds})=\frac{\eta\left(1-\eta+\eta y_{ds}^{\eta}\right)}{2-\eta+ y_{ds}^{\eta}+...}
\ee
which clearly satisfies $0<m(y_{ds})< 1$ provided that $0<\eta<1$.\\
If one assumes that $y_{ds}>>1$, then from (\ref{dsmodel-l}) follows that $\lambda\approx y_{ds}/2$. The following approximation is valid for the deviation parameter $m$ in the $R>>\mu^2$-regime, as seen from Eq.  (\ref{mmodel-l})
\be\label{mapprox1-l}
m(y)\approx \frac{\eta^2 y_{ds} y^{2\eta-1}}{2e^{y^{\eta}}},
\ee
and for the parameter $r$ from (\ref{rmodel-l}), the same expression given by the Eq. (\ref{rapprox1}) is obtained, which allows to write explicitly $m(r)$ as
\be\label{mapprox}
m(r)\approx \frac{1}{2}y_{ds}\eta^2\left(\frac{y_{ds} r}{2(r+1)}\right)^{2\eta-1}e^{-\left(\frac{y_{ds} r}{2(r+1)}\right)^{\eta}},
\ee
which shows a quit different behavior compared to (\ref{mapprox1}) due to the exponential factor that strongly affects the slope of $m$.
To find numerically the background evolution of the model we use (\ref{back2}) for $\mu^2$ together with (\ref{current-l2}) and (\ref{current-l2b}). In Fig. 2 we show some examples  of the evolution of $w_{DE}$ and $\Omega_{DE}$\\

\begin{figure}[h]
\begin{center}
\includegraphics[scale=0.59]{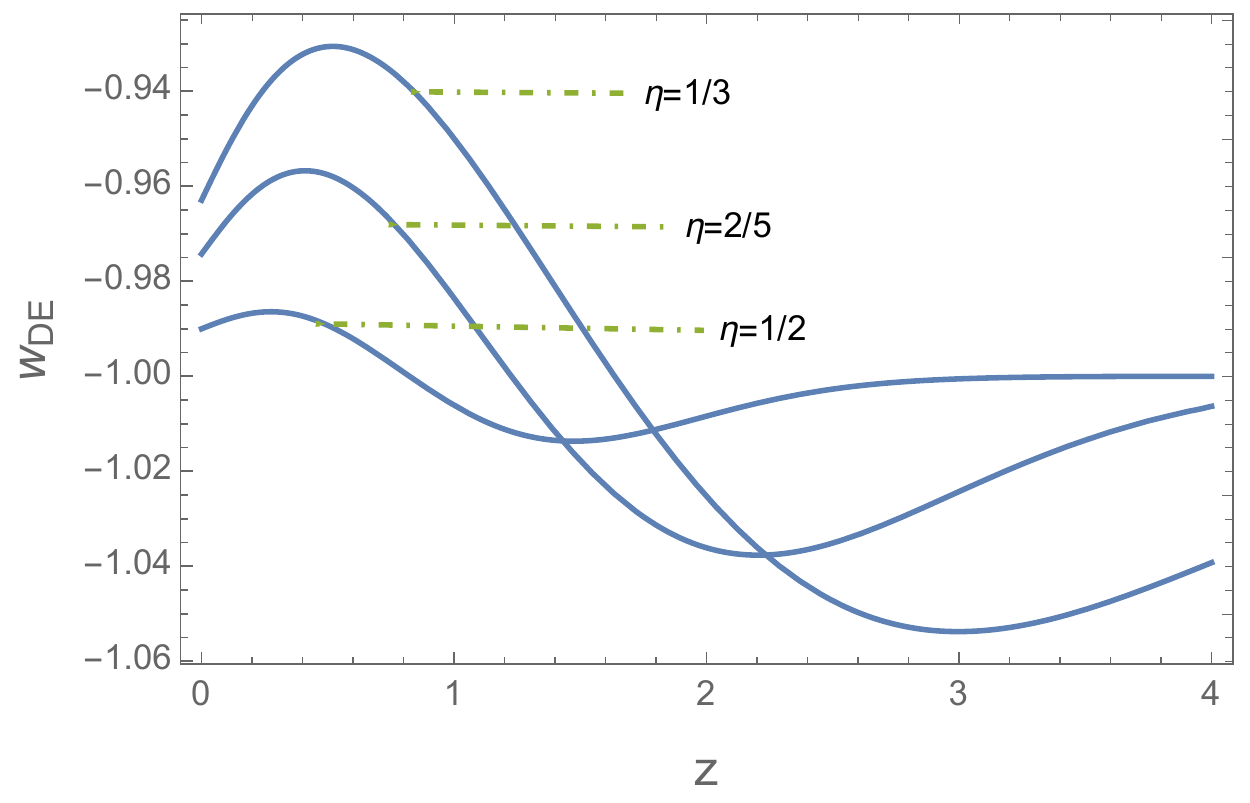}
\includegraphics[scale=0.59]{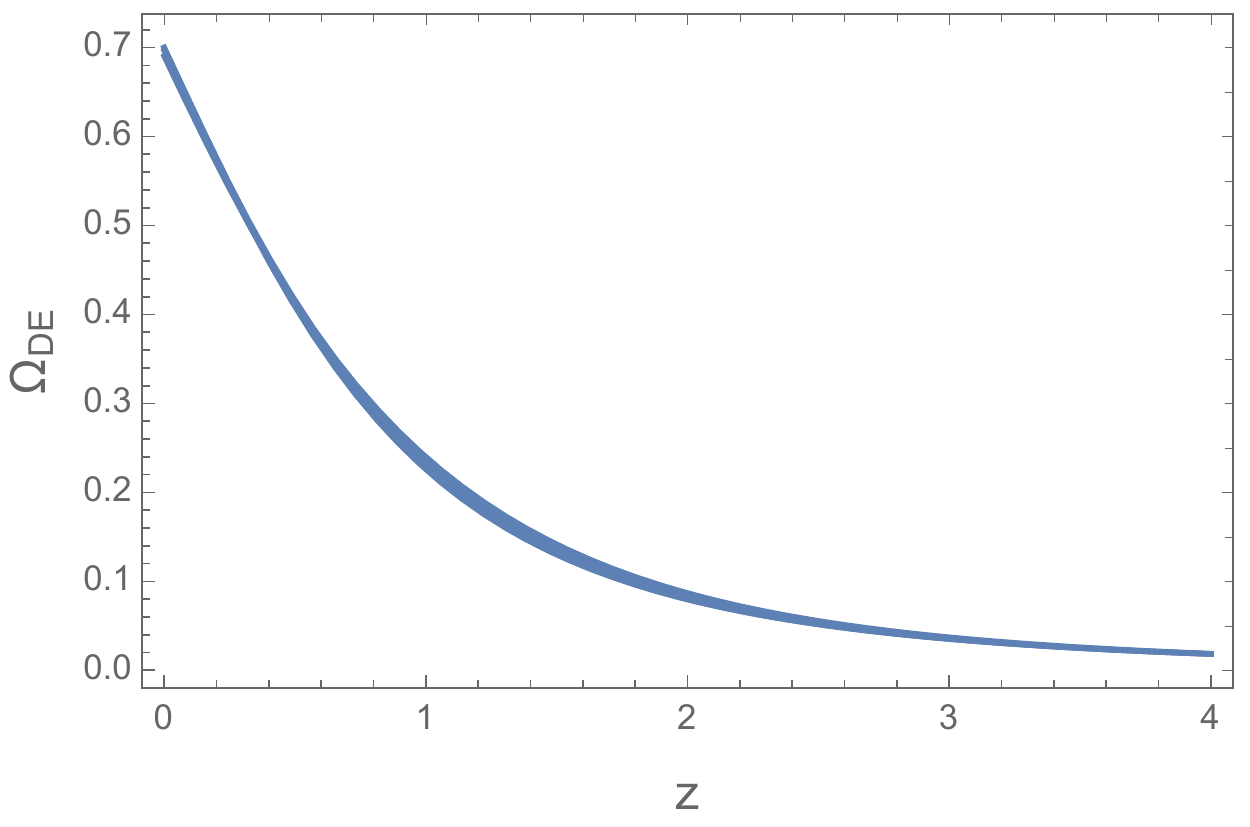}
\caption{The evolution of $w_{DE}$ and $\Omega_{DE}$ for the model (\ref{model-l1}), assuming $\Omega_{m0}=0.3$. The initial redshift is $z_i=6.39$ for al cases. In all cases the evolution of $\Omega_{DE}$ is indistinguishable from that of $\Lambda$CDM. Comparing with Fig. 1 it can be seen that the phantom behavior is more perceptible than in the model (\ref{model2}).}
\label{fig2}
\end{center}
\end{figure}

\noindent {\bf Starobinsky and Hu-Sawicki models}\\
\noindent Other types of viable models are generated by the $g_2(R)$ function
\be\label{model-s}
g_2(R)=\alpha\ln\left[1+\left(\frac{R}{\mu^2}\right)^{\eta}\right]
\ee
with $\alpha>0$, $\eta>0$ . The $f(R)$ function takes the form
\be\label{model-s1}
f(R)=R-\lambda\mu^2\left[1-\left(1+\left(\frac{R}{\mu^2}\right)^{\eta}\right)^{-\alpha}\right],
\ee
The corresponding deviation parameters are given by
\be\label{rcs}
r=-\frac{y-\alpha\eta\lambda y^{\eta}\left(y^{\eta}+1\right)^{-\alpha-1}}{y+\lambda\left(\left(y^{\eta}+1\right)^{-\alpha}-1\right)}
\ee
\be\label{mcs}
m=\frac{\alpha\eta\lambda y^{\eta}\Big[\left(\alpha\eta+1\right)y^{\eta}-\eta+1\Big]}{\left(y^{\eta}+1\right)\Big[y\left(y^{\eta}+1\right)^{\alpha+1}-\alpha\eta\lambda y^{\eta}\Big]}.
\ee
Performing the same analysis as with the previous models, we find the following approximate expression for $m$ in the regime $R>>\mu^2$
\be\label{mmodel-s}
m\approx \frac{\alpha\eta\left(\alpha\eta+1\right)y_{ds}}{2y^{\eta\alpha+1}},
\ee
which in fact is also valid at late times whenever $y_{ds}>>1$.
This expression depends only on the product $\alpha\eta$,  which leads to degeneracy. Then we can find equivalent models (under the regime $\mu^2<<R$) by setting one of the parameters to $1$. Setting  $\eta=1$ in (\ref{model-s1}) we find a model that gives the same results as the Starobinsky model \cite{astarobinsky}. Setting $\alpha=1$ in  (\ref{model-s1}) gives the HS model (with $c_2=1$ in \cite{hu}).
Note also that the function $g_2(R)=\ln\left[1+\lambda_1\left(\frac{R}{\mu^2}\right)^{\eta}\right]$ gives the  HS model.\\
\noindent To estimate the difference between the model (\ref{model2}) and the HS model we can use the almost model-independent initial condition encoded in the amplitude of the cosmological field $\tilde{f}_{,R}=f_{,R}-1$. Then, starting from the same initial condition, the behavior of the models can be followed. If we set the initial condition $|\tilde{f}_{,R}(R_0)|=\tilde{f}_{R_0}$, then we can determine the constants $\alpha$ in (\ref{model2}) and 
$c_2$ in HS, which is described by
\be\label{hs1}
f(R)=R-\frac{c_1\mu^2\left(\frac{R}{\mu^2}\right)^{\eta}}{1+c_2\left(\frac{R}{\mu^2}\right)^{\eta}}.
\ee
From (\ref{back2}), (\ref{1model2}), (\ref{current-l2b}) and (\ref{R0}) we find for the model (\ref{model2})
\be\label{ini2}
\tilde{f}_{R_0}=\frac{6\alpha\eta(1-\Omega_{m0})}{\Omega_{m0}}\left(\frac{12}{\Omega_{m0}}-9\right)^{-\eta-1}\Big[\left(\frac{12}{\Omega_{m0}}-9\right)^{-\eta}+1\Big]^{-\alpha-1},
\ee
which for a given $\eta$ defines $\alpha$ in terms of $\tilde{f}_{R_0}$ 
\be\label{alphaf}
\alpha=-\frac{W\Big[-\frac{\tilde{f}_{R_0}\Omega_{m0}}{6\eta(1-\Omega_{m0})}\left[1+y_0^{\eta}\right]y_0\ln\left[1+y_0^{-\eta}\right]\Big]}{\ln\left[1+y_0^{-\eta}\right]}.
\ee
where $W$ is  the Lambert function and $y_0=\left(\frac{12}{\Omega_{m0}}-9\right)$. Then, replacing $\alpha(\tilde{f}_{R_0})$ in (\ref{mcs}) we find the  current value of the deviation parameter $m_0$.
Likewise, for the model (\ref{hs1}) we find ($\Omega_{\Lambda}=1-\Omega_{m0}$)
\be\label{c3f}
c_2=\frac{y_0^{-2\eta}\left[\eta\Omega_{\Lambda}y_0^{\eta}-(4-3\Omega_{m0})y_0^{\eta}\tilde{f}_{R_0}+\sqrt{\eta\Omega_{\Lambda}y_0^{2\eta}\left(6\Omega_{m0}\tilde{f}_{R_0}-8\tilde{f}_{R_0}+n\Omega_{\Lambda}\right)}\right]}{(4-3\Omega_{m0})\tilde{f}_{R_0}}.
\ee
In Fig. 3 we show the percentage difference between the deviation parameters of models (\ref{model2}) and (\ref{hs1}) for $\eta=4$, at current epoch ($m_0=m(R_0$), for the interval of initial values $0.01\le \tilde{f}_{R_0}\le 0.1$. The growth rate $f$ (see the next section) is also shown for $\tilde{f}_{R_0}=10^{-2}$.
\begin{figure}[h]
\begin{center}
\includegraphics[scale=0.6]{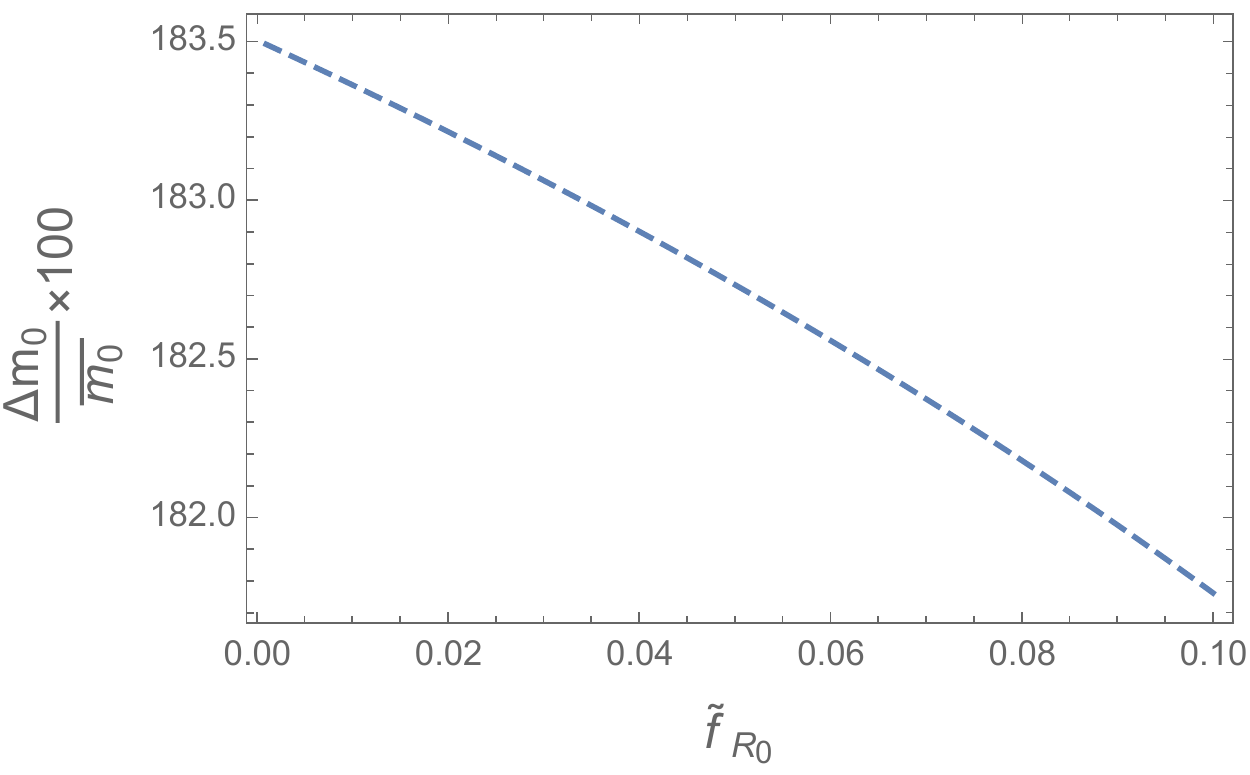}
\includegraphics[scale=0.55]{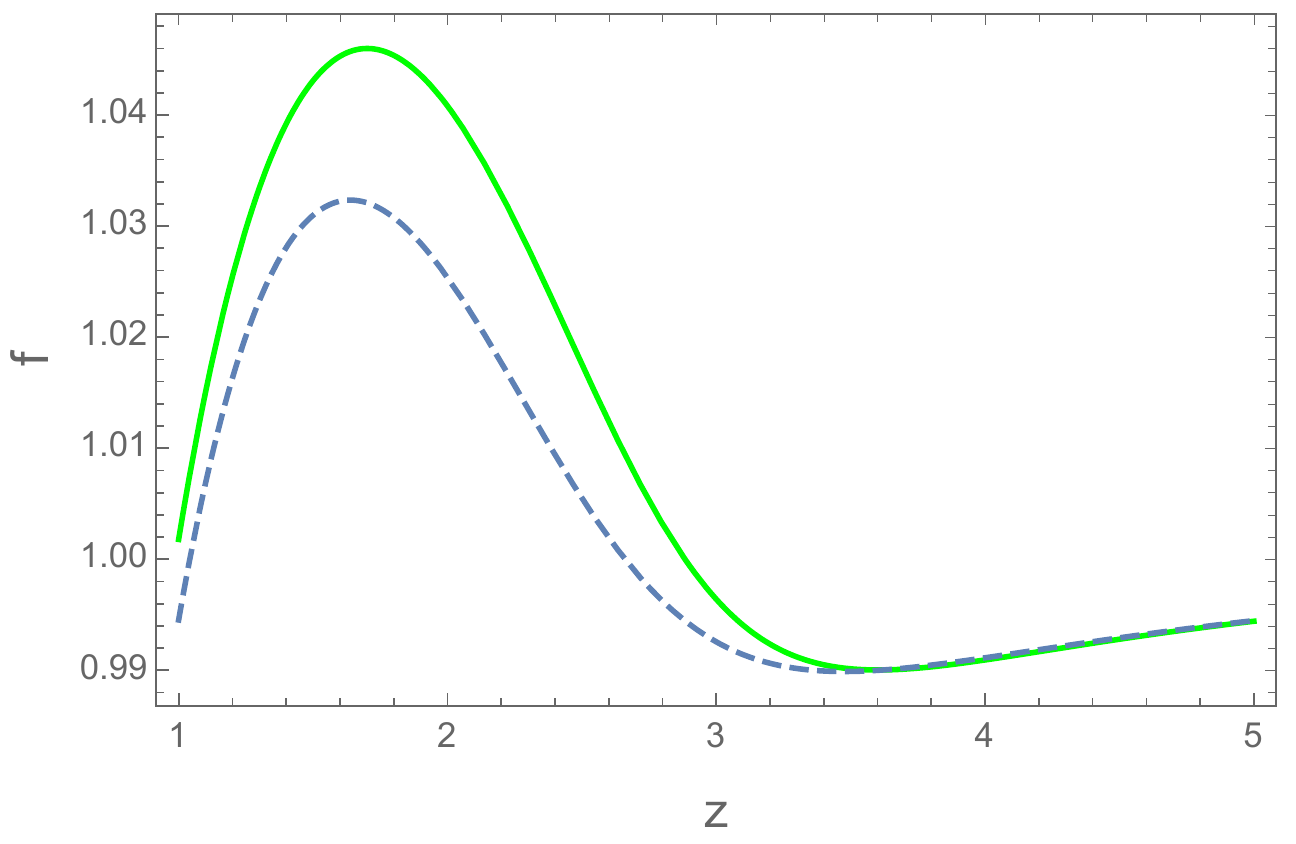}
\caption{\it The percentage deviation for $m_0$ between the models (\ref{model2}) and (\ref{hs1}) for initial values in the interval $0.001\le \tilde{f}_{R_0}\le 0.1$ and $\eta=4$. For $\tilde{f}_{R_0}<10^{-3}$ the percentage deviation tends to settle around $\sim 183.5\%$. This difference in the initial value of $m$ ($m$ varies in the range $10^{-2}$ to $10^{-4}$ along this interval) is reflected in the behavior of the growth rate of matter perturbations as shown in the right curve for the mode $k=600 a_0 H_0$ and $\tilde{f}_{R_0}=10^{-2}$, where the green line corresponds to the model (\ref{model2}) and the dashed line to the model (\ref{hs1}). The curves differ in the maximum $f$ and in the transition redshift. The results have been obtained from the exact formulas for $m$.}
\label{fig3}
\end{center}
\end{figure}
Finding such small differences demands precision in observations of the universe, both late and at high redshift, that is not within the scope of current experiments.
\subsection*{Chameleon Mechanism}
To avoid conflict with local gravity tests, such as the solar system, an important effect named chameleon mechanism \cite{amanda1,amanda2} can be used. This effect is due to the propagation of the scalar field (scalaron) associated to  $f(R)$, whose effective potential is described by the trace equation (\ref{trace}). From the chameleon mechanism follows that the scalar field mass $m_{\phi}$ depends on the local matter density, being large for high-density environments and reducing to smaller values for low-density environments. In the cosmological background, for instance, $m_{\phi}\sim H_0$. In the Solar system this chameleon field gives important information about the strength of the force it mediates and the post-Newtonian parameter $\gamma$, which can be used to test the viability of $f(R)$ gravity models. This scalar field that appears in the Einstein frame after the scale transformation of the metric, with the factor
\be\label{conformal1}
f_{,R}=e^{-\sqrt{\frac{2}{3}}\phi/M_p},
\ee
couples to the matter and gives rise to the scalar field potential  
\be\label{conformal2}
V(\phi)=\frac{M_p^2}{2}\frac{Rf_{,R}-f(R)}{f_{,R}^2}.
\ee
The coupling to the matter Lagrangian gives rise to the effective potential \cite{amanda1, amanda2, tegmark}
\be\label{solar1}
V_{eff}(\phi)=V(\phi)+e^{\beta\phi/M_p}\rho
\ee
where $\rho$ is the matter density in the Einstein frame,  $\beta=1/\sqrt{6}$ is a constant universal coupling between matter and the scalaron $\phi$ that originates in the conformal transformation. 
In the Solar system we consider the Sun as a spherically symmetric object of radius $r_S$ and mass $M_S$ surrounded by background matter at much lower density. We will assume that it has constant density $\rho_S$ (for $r<r_S$) and outside the body ($r>r_S$) the density $\rho_B$ satisfies $\rho_B<<\rho_S$ ($\rho_B\approx 10^{-24} g/cm^2$ is the local homogeneous matter density in our Galaxy). The gravitational potential on the surface of the body is $\Phi_S=GM_S/r_S$, where $M_S=(4/3)\pi r_S^3 \rho_S$. 
Then the effective potential (\ref{solar1}) for the Solar system evolves in two different density environments, presenting two different minima at the field values denoted as $\phi_S$ and $\phi_B$, i.e.
\be\label{solar2}
V'(\phi_S)+\frac{\beta}{M_p}e^{\beta\phi/M_p}\rho_s=0, \;\;\; r<r_S
\ee
\be\label{solar3}
V'(\phi_B)+\frac{\beta}{M_p}e^{\beta\phi/M_p}\rho_B=0, \;\;\; r>r_S,
\ee
where the in high-density region ($r<r_S$) the scalar field acquires the mass $m_S^2=V''_{eff}(\phi_S)$, whereas in the background region $m_B^2=V''_{eff}(\phi_B)$. 
In spherically symmetric spacetime the scalar field obeys the equation of motion (($r$ is the radial distance in spherical coordinates))
\be\label{solar4}
\frac{d^2\phi}{dr^2}+\frac{2}{r}\frac{d\phi}{dr}=\frac{dV_{eff}}{d\phi}
\ee
Solving this equation with the appropriate boundary conditions, it was found in \cite{amanda1,amanda2} that the exterior solution for large bodies like the Sun or the Earth develops a thin shell, giving rise to the following expression for the scalar field  ($r>r_S$)  
\be\label{solar5}
\phi(r)\approx \frac{\beta}{4\pi M_p}\left(\frac{3\Delta r_S}{r_S}\right)\frac{M_S e^{-m_B r}}{r}+\phi_B
\ee
where the thin-shell parameter ($\Delta r_S/r_S$), given by
\be\label{solar6}
\frac{\Delta r_S}{r_S}=\frac{\phi_B-\phi_S}{6\beta M_p\Phi_S},
\ee
must be much smaller than unity in order to suppress the chameleon effect. \\
Applied to the above considered $f(R)$ models, in order to estimate the thin-shell parameters we need to find the corresponding fields $\phi_S$ and $\phi_B$ from the Eqs. (\ref{solar2}) and  (\ref{solar3}). Taking into account that the condition $\mu^2<<R$ follows, the effective potential for the models (\ref{model2}) and (\ref{model-l1}) can be written respectively as (we will use $V^{(1)}_{eff}$ for the model (\ref{model2}) and $V^{(2)}_{eff}$ for the model (\ref{model-l1}) and likewise for the corresponding scalar fields)

\be\label{solar7}
V^{(1)}_{eff}=\frac{1}{2}\lambda \mu^2 M_p^2 e^{2\sqrt{\frac{2}{3}}\phi/M_p}\left[1-\alpha(\eta+1)\left(\sqrt{\frac{2}{3}}\frac{1}{\lambda\alpha\eta}\frac{\phi}{M_p}\right)^{\frac{\eta}{\eta+1}}\right]+e^{\frac{1}{\sqrt{6}}\phi/M_p}\rho,
\ee
\be\label{solar8}
V^{(2)}_{eff}=\frac{M_p^2}{2}\left(\frac{Rf_{,R}(R)-f(R)}{f^2_{,R}(R)}\right)+\rho e^{\beta\phi/M_p}
\ee
Note that in this last case the curvature $R$ cannot be expressed explicitly in terms of the scalar field, but the scalar field can be expressed in terms of $R$ via the conformal transformation  (\ref{conformal1}). Using the approximation $\phi<<M_p$, we find the following solutions for the minima of the potentials (\ref{solar7}) and (\ref{solar8})
\be\label{solar9}
\phi^{(1)}_S=\sqrt{\frac{3}{2}}\eta\lambda\alpha M_p\left(2\lambda+\frac{\rho_S}{\mu^2 M_p^2}\right)^{-(\eta+1)},\;\; r<r_S
\ee
\be\label{solar10}
\phi^{(1)}_B=\sqrt{\frac{3}{2}}\eta\lambda\alpha M_p\left(2\lambda+\frac{\rho_B}{\mu^2 M_p^2}\right)^{-(\eta+1)},\;\; r>r_S
\ee
and 
\be\label{solar11}
\phi^{(2)}_S=\sqrt{\frac{3}{2}}\eta\lambda M_p\left(2\lambda+\frac{\rho_S}{\mu^2 M_p^2}\right)^{\eta-1}e^{-\left(2\lambda+\frac{\rho_S}{\mu^2 M_p^2}\right)^{\eta}},\;\; r<r_S
\ee

\be\label{solar12}
\phi^{(2)}_B=\sqrt{\frac{3}{2}}\eta\lambda M_p\left(2\lambda+\frac{\rho_B}{\mu^2 M_p^2}\right)^{\eta-1}e^{-\left(2\lambda+\frac{\rho_B}{\mu^2 M_p^2}\right)^{\eta}},\;\; r>r_S
\ee
Note that in all above expressions we can neglect $\lambda$ compared to $\rho_{S,B}/(\mu^2M_p^2)$ ($2\lambda\sim y_{ds}\sim 28$, while, for instance, $\rho_B/(M_p^2\mu^2)\sim \rho_B/\rho_0\sim 10^5$). It can also be observed that the denser the region, the smaller the scalar field. Hence as $\rho_S>>\rho_B$ then $\phi^{(1,2)}_B>>\phi^{(1,2)}_S$, which leads to the following approximation for the thin-shell parameters from (\ref{solar6})
\be\label{solar13}
\frac{\Delta r^{(1)}_S}{r_S}\approx \frac{4\pi M_p^2 r_S}{M_S}\eta\lambda\alpha\left(\frac{\mu^2 M_p^2}{\rho_B}\right)^{\eta+1},
\ee
\be\label{solar14}
\frac{\Delta r^{(2)}_S}{r_S}\approx \frac{4\pi M_p^2 r_S}{M_S}\eta\lambda\left(\frac{\rho_B}{\mu^2M_p^2}\right)^{\eta-1}e^{-\left(\frac{\rho_B}{\mu^2M_p^2}\right)^{\eta}}.
\ee
The result (\ref{solar13}) for the model (\ref{model2}) is similar to the results obtained for the HS and Starobinsky models \cite{tsujikawa}, which applies to models whose deviation parameter $m$ can be expressed in terms of the parameter $r$ (defined in (\ref{r-m})) approximately in the form $m\sim(-1-r)^p$ ($p>1$), provided that $R>>\mu^2$. The behavior of the thin shell parameter (\ref{solar14}) for the model (\ref{model-l1}) is quite different due to the exponential damping which, depending on $0<\eta<1$, can greatly decrease the thickness of the shell. 
The experimental tests of the post-Newtonian parameter $\gamma$ in the Solar system, that give the constraint $|\gamma-1|<2.3\times 10^{-5}$ \cite{cmwill}, allows to find a bound on the thin shell parameter. To this end we use the condition $\phi<<M_p$, what gives rise to the approximation $r=r_{EF}\approx r_{JF}$, where $r_{EF}$ ($r_{JF}$) is the radial distance in the Einstein (Jordan) frame. Then it was found in \cite{tegmark} that under the chameleon mechanism the spherically symmetric solution in the JF can be written as
\be\label{solar15}
ds^2=-\Big[1-\frac{2GM_S}{r}\left(1+\frac{\Delta r_S}{r_S}\right)\Big]dt^2+\Big[1+\frac{2GM_S}{r}\left(1-\frac{\Delta r_S}{r_S}\right)\Big]dr^2+r^2 d\Omega^2,
\ee
where the condition $\lambda_B\sim m_B^{-1}>>r_S$ was used. Then  the post-Newtonian parameter $\gamma$ can be approximated as
\be\label{solar16}
\gamma\approx \frac{1-\Delta r_S/r_S}{1+\Delta r_S/r_S}.
\ee
From this expression and using the experimental restriction on $\gamma$ \cite{cmwill} it is found 
\be\label{solar17}
\frac{\Delta r_S}{r_S}<10^{-5}
\ee
Then using the result (\ref{solar13}) we find the following restriction for the model (\ref{model2})
\be\label{solar18}
\frac{1}{4}\eta\alpha y_{ds}\left(\frac{\mu^2 M_p^2}{\rho_B}\right)^{\eta+1}<10^{-11},
\ee
and for the model  (\ref{model-l1}), from (\ref{solar14}) we find the restriction
\be\label{solar19}
\frac{1}{4}\eta y_{ds}\left(\frac{\rho_B}{\mu^2M_p^2}\right)^{\eta-1}e^{-\left(\frac{\rho_B}{\mu^2M_p^2}\right)^{\eta}}<10^{-11},
\ee
where we used  (\ref{lambda2a-ds}) for $\lambda$ and the value $\Phi_S\sim 10^{-6}$ for the Sun. The following numerical cases demonstrate that the models satisfy the bound (\ref{solar17}) (assuming (\ref{back2}) for $\mu^2$, which gives $\mu^2M_p^2/\rho_B\approx 3\times 10^{-6}$). For the model (\ref{model2}): ($\eta=3,\alpha=10$) gives $\frac{\Delta^{(1)} r_S}{r_S}\approx 1.7\times 10^{-14}$, ($\eta=5,\alpha=10^4$) gives $\frac{\Delta^{(1)} r_S}{r_S}\approx 2.5\times 10^{-22}$. For the model  (\ref{model-l1}): $\eta=1/3$ gives $\frac{\Delta^{(2)} r_S}{r_S}\approx 3.7\times 10^{-28}$, $\eta=2/5$ gives $\frac{\Delta^{(2)} r_S}{r_S}\approx 6.9\times 10^{-68}$. To translate the constraint (\ref{solar18}) to $\eta$ we express $\alpha$ through $\tilde{f}_{R_0}$ using (\ref{alphaf}) and then solve the resulting expression with respect to $\eta$. Thus, for the initial value $\tilde{f}_{R_0}=10^{-2}$ we find $\eta>1.16$ and $\eta<4.4\times 10^{-7}$. This last bound can be ruled out since the model (\ref{model2}) becomes indistinguishable from $\Lambda$CDM. For the model  (\ref{model-l1}) we find from (\ref{solar19}) $\eta>0.17$ and $\eta<1.1\times 10^{-5}$, where the latter can be discarded for the same reasons as above.\\
The chameleon mechanism applied to the Earth leads to important restrictions that also allow to avoid possible violations of the equivalence principle \cite{amanda1,amanda2}. To estimate the thin-shell conditions for the Earth, it is considering the Earth as a solid sphere of radius $r_E\approx 6\times 10^3\;km$ and homogeneous density $\rho_E\approx10\;gr/cm^3$.
First we note that there are two environments surrounding the earth (ignoring the influence of the Sun, the Moon and the other planets): the atmosphere which is approximated as a $10\;km$-thick layer with density $\rho_{atm}\sim 10^{-3}\;g/cm^3$ and the homogeneously distributed matter in our Galaxy with density $\rho_B$. The gravitational potentials $\Phi_E=\rho_Er_E^2/(6M_p^2)$ and $\Phi_{atm}=\rho_{atm}r_{atm}^2/(6M_p^2)$ are related as $\Phi_{atm}\approx 10^{-4}\phi_E$ ($r_{atm}\approx r_E$).
In order for the atmosphere to have a thin-shell, it thickness should be less than $10\;km$, i.e. $\Delta r_{atm}/r_{atm}\approx \Delta r_{atm}/r_E<10^{-3}$. Following the same guidelines as for the Solar system, and denoting the values that minimize $V_{eff}$ as $\phi_E, \phi_{atm}, \phi_B$ for $\rho_E, \rho_{atm}$ and $\rho_B$ respectively, we find 
$\Delta r_E/r_E=(\phi_B-\phi_E)/(\sqrt{6}M_p\Phi_E)$ and $\Delta r_{atm}/r_{atm}=(\phi_{B}-\phi_{atm})/(\sqrt{6}M_p\Phi_{atm})$. Then by analogy with (\ref{solar9}), (\ref{solar10}) we have that $\phi_E<<\phi_B$ and $\phi_{atm}<<\phi_B$, which gives the following upper bound 
\be\label{solar20}
\frac{\Delta r_E}{r_E}\approx \frac{\Phi_{atm}}{\Phi_E}\frac{\Delta r_{atm}}{r_{atm}}<10^{-7}.
\ee 
Which taking into account that $\Phi_E\approx 10^{-9}$ results in the following restrictions for the models (\ref{model2}) and (\ref{model-l1}) respectively
\be\label{solar21}
\frac{1}{4}\eta\alpha y_{ds}\left(\frac{\mu^2 M_p^2}{\rho_B}\right)^{\eta+1}<10^{-16},
\ee
and
\be\label{solar22}
\frac{1}{4}\eta y_{ds}\left(\frac{\rho_B}{\mu^2M_p^2}\right)^{\eta-1}e^{-\left(\frac{\rho_B}{\mu^2M_p^2}\right)^{\eta}}<10^{-16}.
\ee
Compared with the Solar system, the thin-shell restrictions on the Earth are stronger than those imposed by the value of the pos-Newtonian parameter $\gamma$. Assuming $\tilde{f}_{R_0}=10^{-2}$ for $\alpha$ in (\ref{alphaf}) we find from (\ref{solar21}) $\eta>2.4$ or $\eta<1.3\times 10^{-13}$. For the model (\ref{model-l1}) we find from (\ref{solar22}) $\eta>0.22$ or $\eta<1.1\times 10^{-10}$. In both models the last bound can be disregarded since they become indistinguishable from $\Lambda$CDM model.
\section{Restrictions from matter density perturbations}

The Mere cosmic expansion history does not provide sufficient accuracy to distinguish a viable $f(R)$ model from the $\Lambda$CDM model. 
The evolution of linear perturbations can lead to strong observational deviations from $\Lambda$CDM, which are present in the large-scale structure and Cosmic Microwave Background.
In the sub-horizon approximation, deep inside the Hubble radius ($k^2>>a^2H^2$) and using the quasi-static approximation, the evolution of mater perturbations during the matter dominance is controlled by the equation \cite{tsujikawa10, tsujikawa11, tsujikawa01}
\be\label{matter-pert}
\ddot{\delta}_m+2H\dot{\delta}_m-4\pi G_{eff}\rho_m\delta_m\simeq 0
\ee
where $\delta_m\simeq \delta\rho_m/\rho_m$ and $G_{eff}$ is the effective gravitational coupling 
\be\label{newton}
G_{eff}=\frac{G}{f_{,R}}\left[\frac{1+\frac{4k^2f_{,RR}}{a^2f_{,R}}}{1+\frac{3k^2f_{,RR}}{a^2f_{,R}}}\right] =\frac{G}{f_{,R}}\left[\frac{1+\frac{4k^2 m}{a^2 R}}{1+\frac{3k^2 m}{a^2 R}}\right]\simeq\frac{G}{f_{,R}}\left[\frac{4+3M^2 a^2/k^2}{3+3M^2a^2/k^2}\right],
\ee
where $M$ in given by (\ref{mass3}).
Note that the variation of the effective Newtonian coupling affects the expansion rate, which depends on G, and is critical for the  process of  the Big Bang nucleosynthesis (BBN) since it affects the time of production of primordial light elements, and therefore its abundance  \cite{craig, turner, serpico}. This impose limits on possible variations in Newton constant G. Then, by the time of BBN the evolution of matter perturbations in $f(R)$ gravity must be very similar to that of the GR, which places constraints on $f(R)$ parameters \cite{scarpetta, neseris} via constraints on $G_{eff}$. From (\ref{newton}), this constraint can be expressed as $\xi=\frac{k^2 m}{a^2 R}<<1$.\\
Numerical analysis for the models (\ref{model2}) and (\ref{model-l1}), using the respective expressions for $f_{,R}$ and (\ref{maproxmodel2}) and \ref{mapprox1-l} for $m$ (see also (\ref{mapprox2}) and (\ref{mapprox3}) bellow), and using the $k$-mode corresponding to the horizon size at BBN, $\lambda_{hor}\sim 10^{-4} Mpc$, ($k_{BBN}\sim a_{BBN}/\lambda_{hor}\sim 10^{-5} Mpc^{-1}$ ($a_{BBN}\sim 10^{-9}a_0$), shows that the restriction $\xi<<1$ is generously satisfied, given that the models satisfy the thin shell conditions. For $\tilde{f}_{R_0}=10^{-2}$ in (\ref{alphaf}) and $\eta=2.4$ for the model (\ref{maproxmodel2}) we find $\xi\sim 10^{-99}$ and taking $\eta=0.22$ for the model \ref{mapprox1-l} gives $\xi\sim 10^{-400000}$. Furthermore, as $\eta$ increases in both models these values become even lower. Therefore, given that the models satisfy the thin shell conditions, the BBN does not impose additional restrictions. In fact the bound $\xi<<1$ maintains up to deep matter dominated era where the transition to scalar-tensor regime may occur as shown bellow.\\
During matter dominance, the deviation parameter $m$ for viable $f(R)$ models must satisfy $m<<1$, which is equivalent to a large mass $M$ according to Eqs. (\ref{mass1}) and (\ref{mass3}). The region $M^2>>k^2/a^2$ corresponds to the GR regime ($G_{eff}\simeq G/f_{,R}\simeq G$), where $\delta_m$ evolves as $\delta m\propto t^{2/3}$ during the matter dominance. At latter times the perturbations can enter the scalar-tensor regime that takes place for $M^2<<k^2/a^2$ with the effective gravitational coupling $G_{eff}\simeq 4G/(3f_{,R})\simeq 4G/3$, and the evolution of $\delta_m$ is different, behaving as $\delta_m\propto t^{(\sqrt{33}-1)/6}$  \cite{tsujikawa10, tsujikawa11, tsujikawa01}. From (\ref{newton}) we can see that the transition from GR regime to the modified gravity or scalar-tensor regime occurs when
\be\label{transition}
m= \frac{a^2R}{k^2}
\ee
For wave numbers in the interval 
\be\label{k-linear}
0.01h Mpc^{-1}\lesssim k\lesssim 0.2 h Mpc^{-1},\;\;\; h=0.72\pm 0.08,
\ee
relevant to the linear regime of the galaxy power spectrum \cite{tegmark1, tegmark2, linder, percival, huterer, tsujikawa3}, if the transition to scalar-tensor regime occurred in the current epoch ($z\approx 0$) for the upper bound $k\approx 0.2h$ Mpc$^{-1}\approx 600 a_0H_0$, then according to (\ref{transition}) the deviation parameter is constrained to values
\be\label{m-today}
m(z\approx 0)\gtrsim 3\times 10^{-6},
\ee
which would be within reach of observations in the near future. If the transition occurs during the deep matter era, then the redshift $z_k$ for the transition point can be estimated for a given model if the deviation parameter $m$ as function of $R$ is known. For the models (\ref{model2}) and (\ref{moldel-l}) the expressions for $m$ during matter dominance were found and are given by the Eqs. (\ref{maproxmodel2}) and \ref{mapprox1-l} respectively. On the other hand, according to the results in Figs. 1 and 2, one can assume the $\Lambda$CDM model for the background expansion, and use the following expression for $R$
\be\label{matter-dom}
R\simeq 3H_0^2\left[\Omega_{m0}(1+z)^3+4\Omega_{\Lambda}\right],
\ee
which allows us to find from (\ref{transition}) the transition redshift $z_k$\\
{\bf Model (\ref{model2})}\\
\be\label{mapprox2}
m\approx \frac{1}{2}\alpha\eta(\eta+1)y_{ds} \left(\frac{\mu^2}{R}\right)^{\eta+1}
\ee
\be\label{ztransition1}
(1+z_k)^{-2}\left[\Omega_{m0}(1+z_k)^3+4\Omega_{\Lambda}\right]^{\eta+2}=\frac{1}{6}\alpha\eta(\eta+1)y_{ds}\left(\frac{\Omega_{m0}}{ 3}\right)^{\eta+1}\left(\frac{k}{a_0H_0}\right)^2
\ee

{\bf Model (\ref{moldel-l})}\\
\be\label{mapprox3}
m \approx \frac{\eta^2 y_{ds}}{2}  \left(\frac{R}{\mu^2}\right)^{2\eta-1}\exp\left[- \left(\frac{R}{\mu^2}\right)^{\eta}\right]
\ee
 \be\label{ztransition2}
\frac{\exp\Big[\Big(\frac{3}{\Omega_{m0}}\left[\Omega_{m0}(1+z_k)^3+4\Omega_{\Lambda}\right]\Big)^{\eta}\Big]}{(1+z_k)^2\left[\Omega_{m0}(1+z_k)^3+4\Omega_{\Lambda}\right]^{2\eta-2}}=\frac{1}{6}\eta^2 y_{ds}\left(\frac{3}{\Omega_{m0}}\right)^{2\eta-1}\left(\frac{k}{a_0H_0}\right)^2
 \ee
where we used the approximation for $m$ valid during matter dominated epoch ($\mu^2<<R$) $\mu^2=\Omega_{m0}H_0^2$ (see (\ref{back2})) and the conditions (\ref{lambda2a-ds}) and (\ref{current-l2}) valid for both models. Some values of $m(z\approx 0)$ and $z_k$ are listed in tables I and II. 
\begin{center}
\begin{tabular}{|l|c|c|c|}\hline
\multicolumn{4}{|c|}{\bf Model (\ref{model2})}\\ \hline\hline
 $\eta$  \hspace{0.6cm} & $\alpha$  \hspace{0.6cm}  & $m(z\approx 0)$ \hspace{0.6cm}  &$z_k$ ($\frac{k}{a_0H_0}= 300$) \\ \hline 
 $3$ & $10$ &  $6.6\times 10^{-4}$ & $1.96$ \\ 
 $4$ & $10^2$ & $2.7\times 10^{-4}$ & $1.47$\\ 
 $5$ & $10^4$ &  $10^{-3}$ & $1.63$\\ 
$6$ & $10^6$  & $3.6\times 10^{-3}$ & $1.73$\\ 
 \hline
\end{tabular}
\end{center}
\begin{center}
{{\bf Table I.} \it Some numerical values for the transition redshift and $m=\frac{Rf_{,RR}}{f_{,R}}$ at current epoch ($m(z=0)$), where we have used $\Omega_{m0}\approx 0.3$. Note that with the increase of $\eta$ to the next integer, $\alpha$ must be increased by orders of magnitude, in order to satisfy the bound $m(z\approx 0)>10^{-6}$.}
\end{center}

\begin{center}
\begin{tabular}{|l|c|c|c|}\hline
\multicolumn{3}{|c|}{\bf Model (\ref{moldel-l})}\\ \hline\hline
 $\eta$  \hspace{0.6cm}  & $m(z\approx 0)$ \hspace{0.6cm}  &$z_k$ ($\frac{k}{a_0H_0}= 300$) \\ \hline 
 $1/4$ &  $0.0288$ & $7.48$ \\ 
 $1/3$ &  $0.0258$ & $5.16$\\ 
 $2/5$ &  $0.0190$ & $3.82$\\ 
$1/2$ &   $7.4\times 10^{-3}$ & $2.48$\\ 
$2/3$ &  $1.8\times 10^{-4}$ & $1.06$\\ 
 \hline
\end{tabular}
\end{center}
\begin{center}
{{\bf Table II.} \it The transition redshift and current values of $m$ for some cases of the model (\ref{moldel-l}). }
\end{center}

\noindent The scalar-tensor regime starts later for larger $\eta$. Note that for $z_k$ smaller than the order of unity, the Universe does not enter the scalar-tensor regime during the matter dominated epoch. 

\noindent {\bf The Growth of Matter Perturbations.}\\

\noindent  
The growth of large scale structure in the universe provides an important test which can reveal a deviation from the $\Lambda$CDM model especially at late times. 
The Eq. (\ref{matter-pert}) for the fractional matter density perturbation $\delta_m$ can be written in terms of the $e$-fold variable $N=\ln a$ as follows
\be\label{growth}
\frac{df(a)}{dN}+f(a)^2+\frac{1}{2}\left(1-\frac{d\ln \Omega_m(a)}{dN}\right)f(a)=\frac{3}{2}\frac{G_{eff}}{G}\Omega_m(a)
\ee
where 
\be\label{growth1}
f(a)=\frac{d\ln\delta_m}{dN},
\ee
is the growth rate and $G_{eff}$ is given in (\ref{newton}) which carries a scale-dependence, and $\Omega_m(a)$ can be read off from Eq. (\ref{eq2a}) by rewriting it in the form
\be\label{eq2a1}
H^2=\frac{\kappa^2}{3}\left(\rho_m+\rho_{DE}\right),\;\; \rho_{DE}=\frac{1}{\kappa^2}\left[\frac{1}{2}\left(RF-f\right)-3H\dot{F}+3H^2\left(1-F\right)\right].
\ee
Then
\be\label{growth2}
\Omega_m(a)=\frac{\kappa^2\rho_m}{3H^2}=\frac{\rho_m}{\rho_m+\rho_{DE}}=\frac{\Omega_{m0}a^{-3}}{H^2/H_0^2},\;\;\; \Omega_{m0}=\frac{\kappa^3\rho_{m0}}{3H_0^2},
\ee
where we have neglected the radiation. Note also that for viable $f(R)$ models at high redshift, during matter dominated epoch, $F\approx 1$ and $\Omega_m$ defined in (\ref{densities}) reduces to the above standard definition.
A widely used representation for $f$ is of the form
\be\label{growth3}
f(a)=\Omega_m(a)^{\gamma(a)},
\ee
where $\gamma$ defined by
\be\label{growth4}
\gamma(a)=\frac{\ln f(a)}{\ln \Omega(a)}
\ee
is the growth index of matter perturbations \cite{peebles,wangstein,linder1}. 
In order to integrate the eq. (\ref{growth}) in the matter dominated epoch we use the fact that in the high redshift region the model (\ref{model2}) is close to the $\Lambda$CDM model, and therefore we can assume that the background expansion is well approximated by the $\Lambda$CDM model. 
Using (\ref{mapprox2}) in $G_{eff}$ given in (\ref{newton}), we show in Fig. 4 the evolution of the growth function $f$  and the growth index $\gamma$ for the modes $k/(a_0H_0)=30, 100, 300, 600$, for two cases of the model (\ref{model2}).
\begin{figure}
\begin{center}
\includegraphics[scale=0.57]{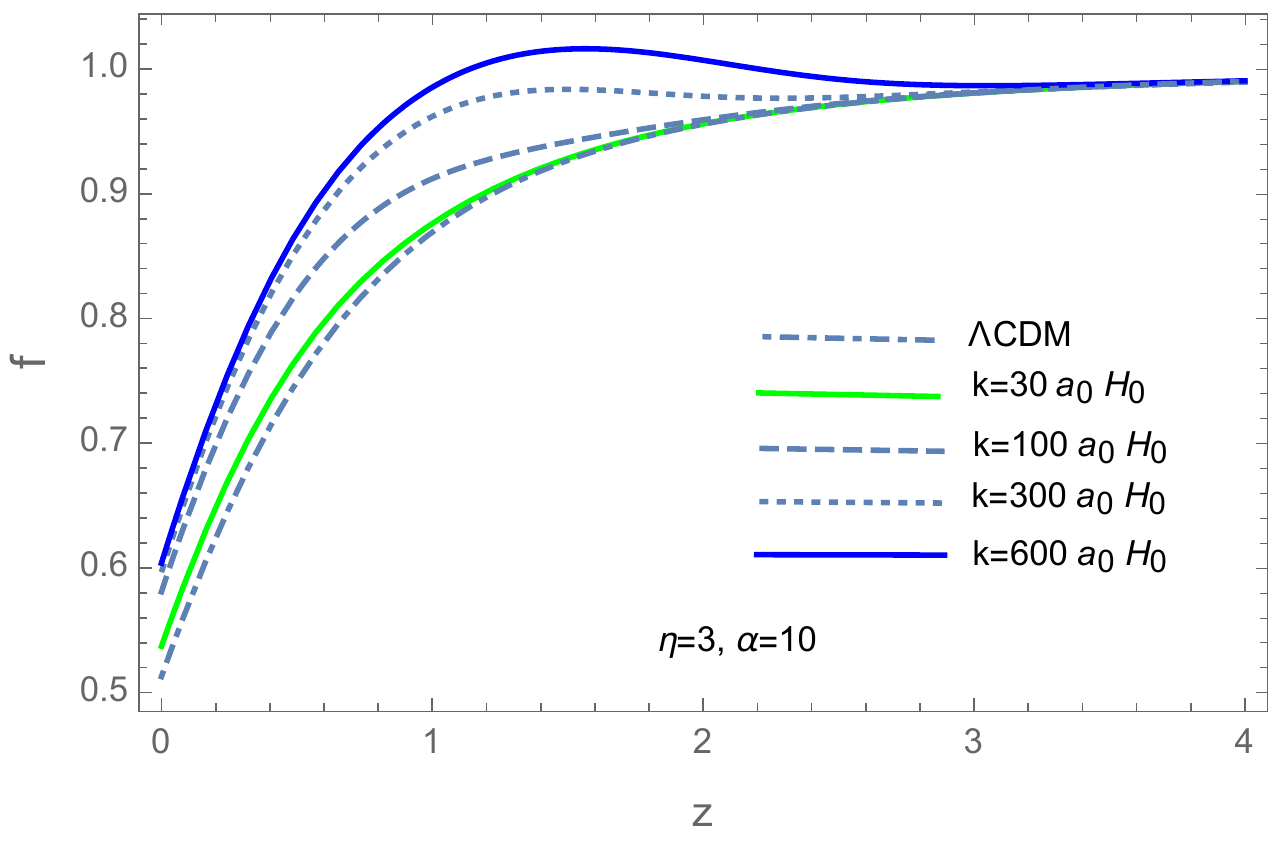}
\includegraphics[scale=0.57]{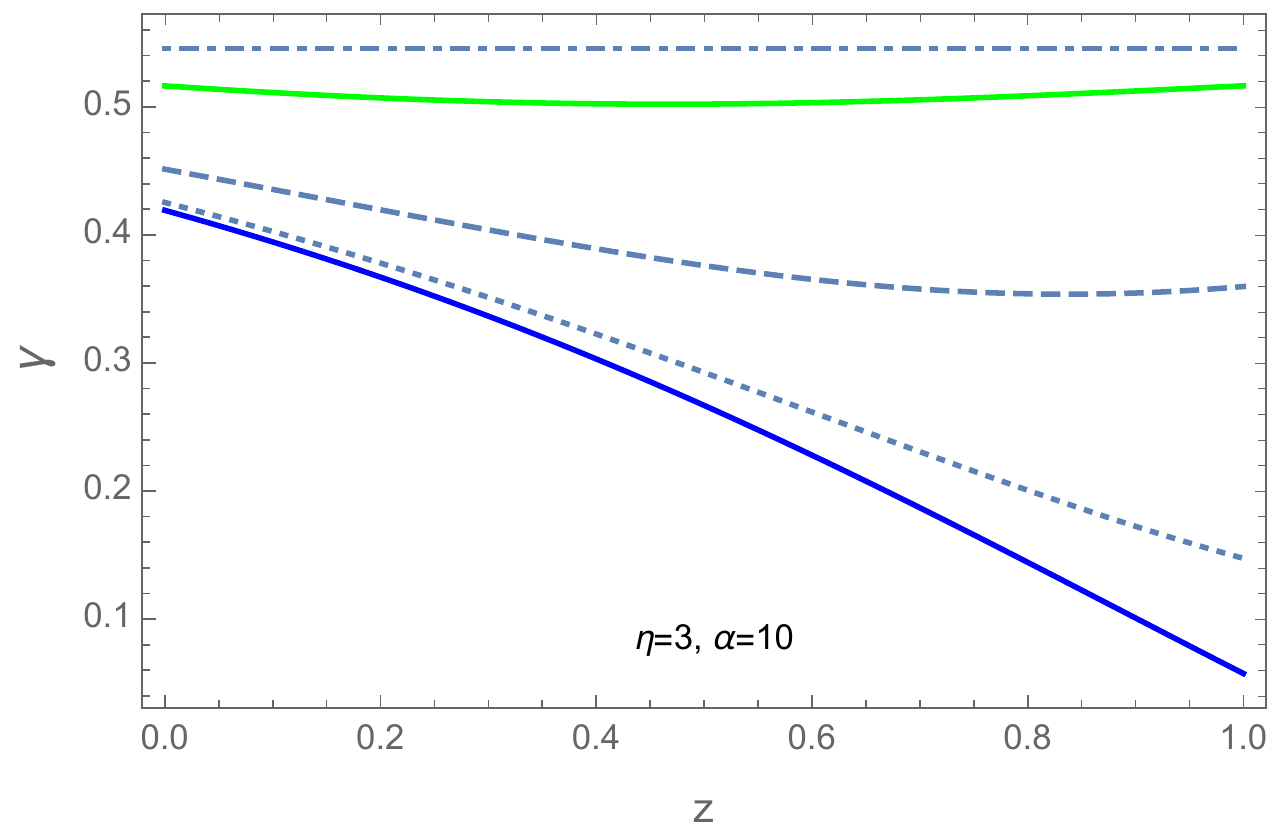}
\includegraphics[scale=0.57]{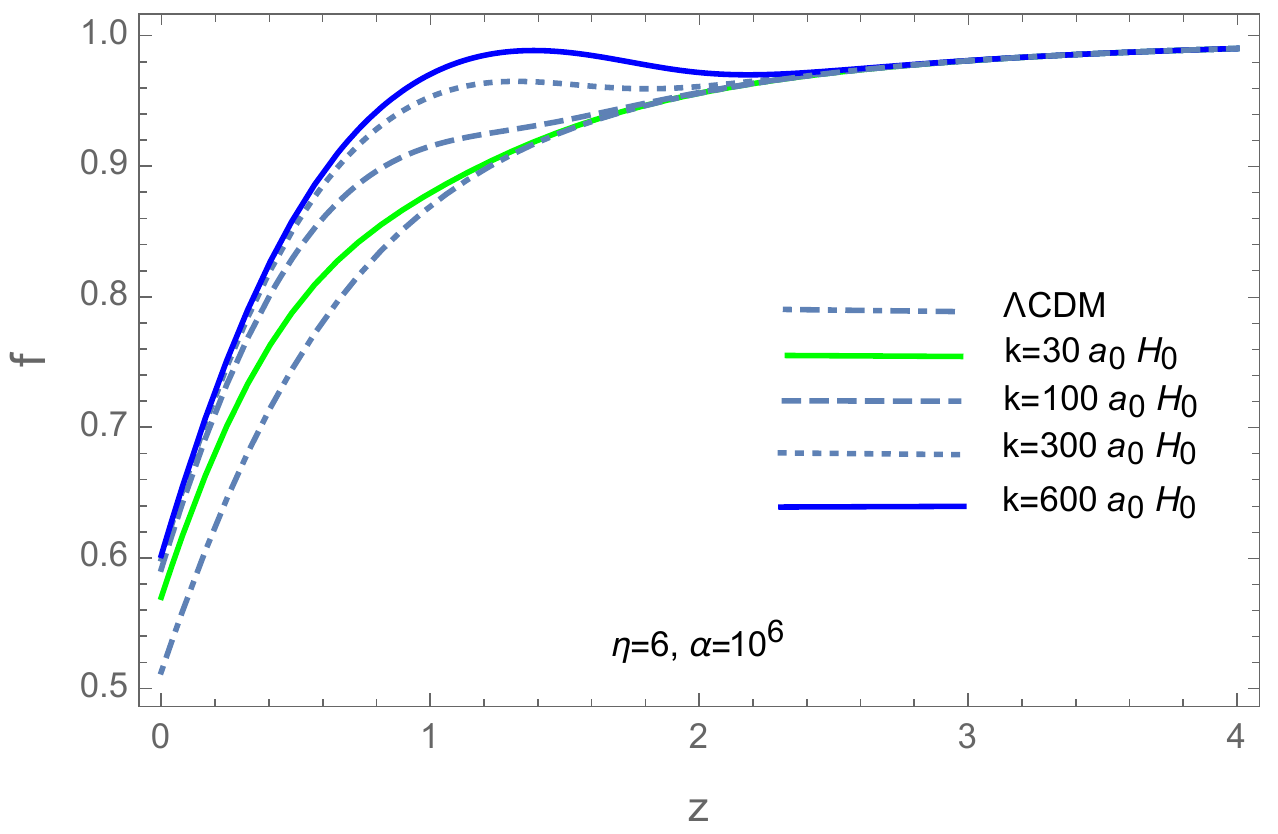}
\includegraphics[scale=0.57]{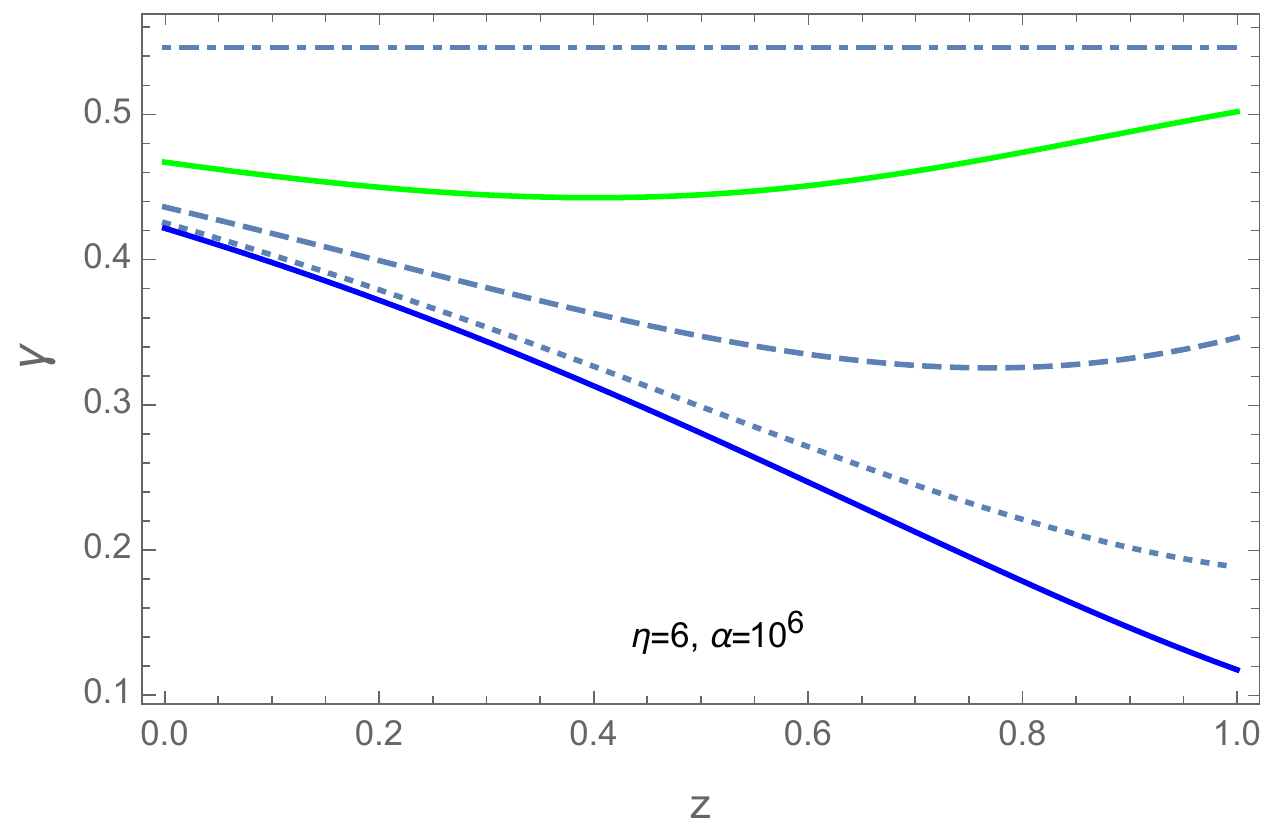}
\caption{The evolution of the growth rate $f$ and the growth index $\gamma$ in the model (\ref{model2}) with $\mu^2=\Omega_{m0}H_0^2$, $\Omega_m=0.3$, $\Omega_{\Lambda}=0.7$. The upper figures correspond to $\eta=3$, $\alpha=10$ and those below to $\eta=6$, $\alpha=10^6$.
The dispersion of $\gamma$ is present for larger scales ($k=30 a_0H_0$ and $k=100 a_0H_0$) at current epoch, indicating that the asymptotic regime $k>>aM$ has not been reached. The transition redshift is larger for smaller $\eta$. For $\eta=3$, $f$ and $\gamma$ evolve very close to $\Lambda$CDM for the larger scale, $k=30 a_0H_0$.}
\label{fig4}
\end{center}
\end{figure}
\noindent The results of Fig.4 show that $f$ and $\gamma$ for the larger modes have reached the scale invariance at current epoch. It can be seen that the maximum values of the growth rate increases with increasing $k$. This suggests that the matter power spectrum is enhanced on smaller scales, which  results in different spectral indices on different scales. Once the Universe enters the late-time accelerated epoch the growth rate begins to decrease.\\
In Fig. 5 we show the evolution of the growth function $f$  and the growth index $\gamma$ for the modes $k/(a_0H_0)=30, 100, 300, 600$, for two cases of the model (\ref{model-l1}) using (\ref{mapprox3}) in (\ref{growth}).
 \begin{figure}[h]
\begin{center}
\includegraphics[scale=0.59]{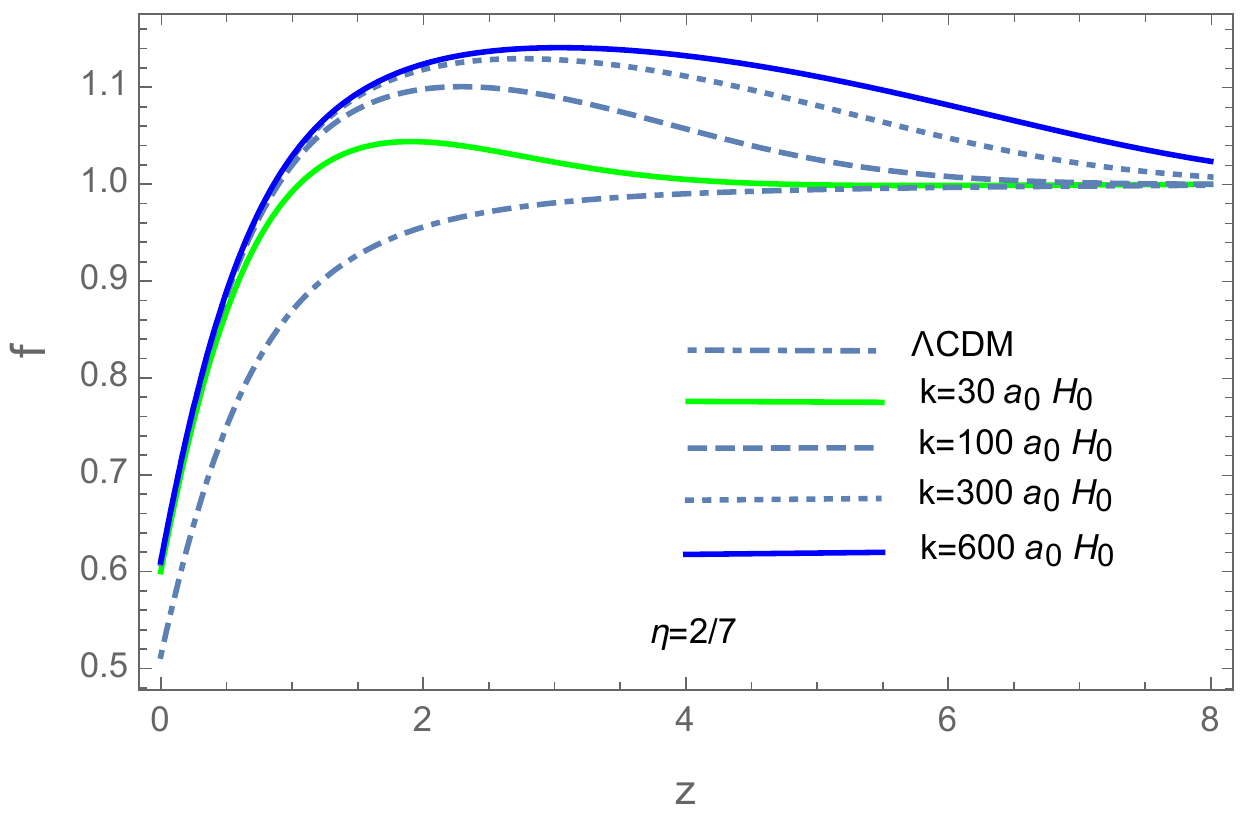}
\includegraphics[scale=0.59]{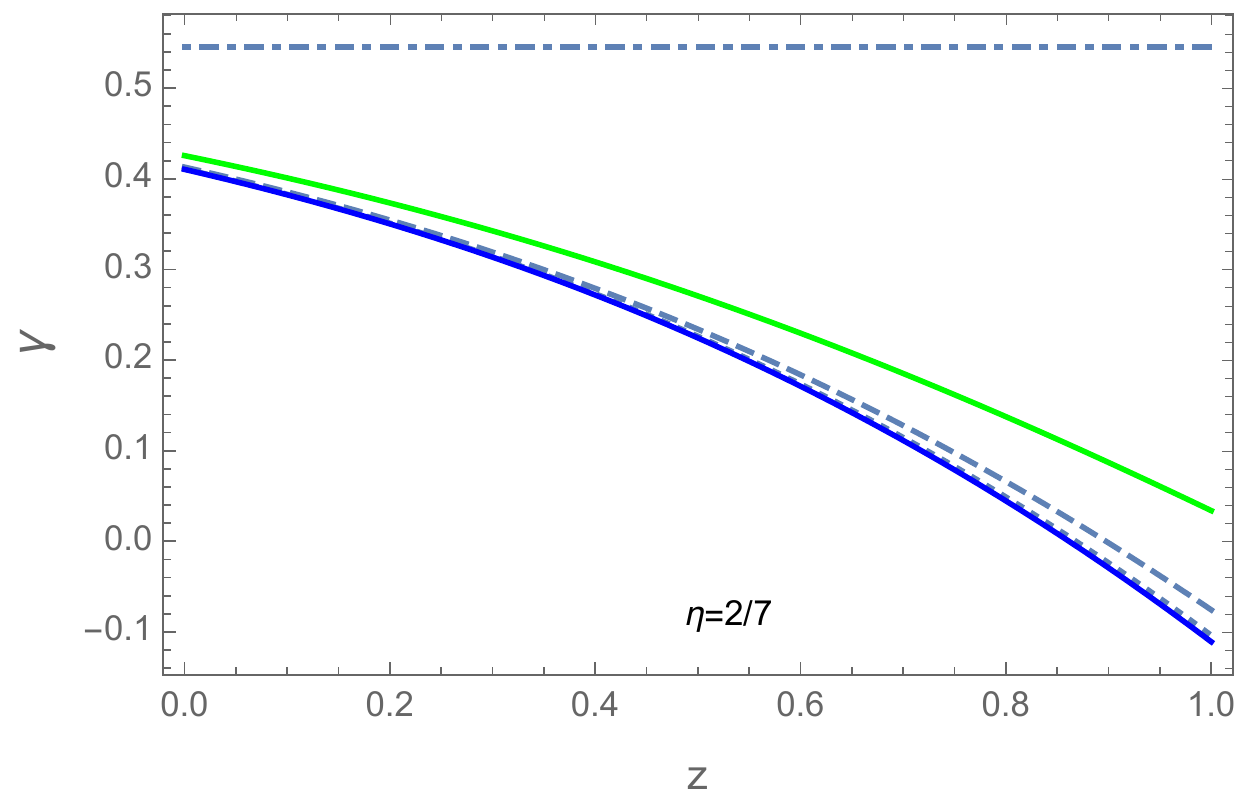}
\includegraphics[scale=0.59]{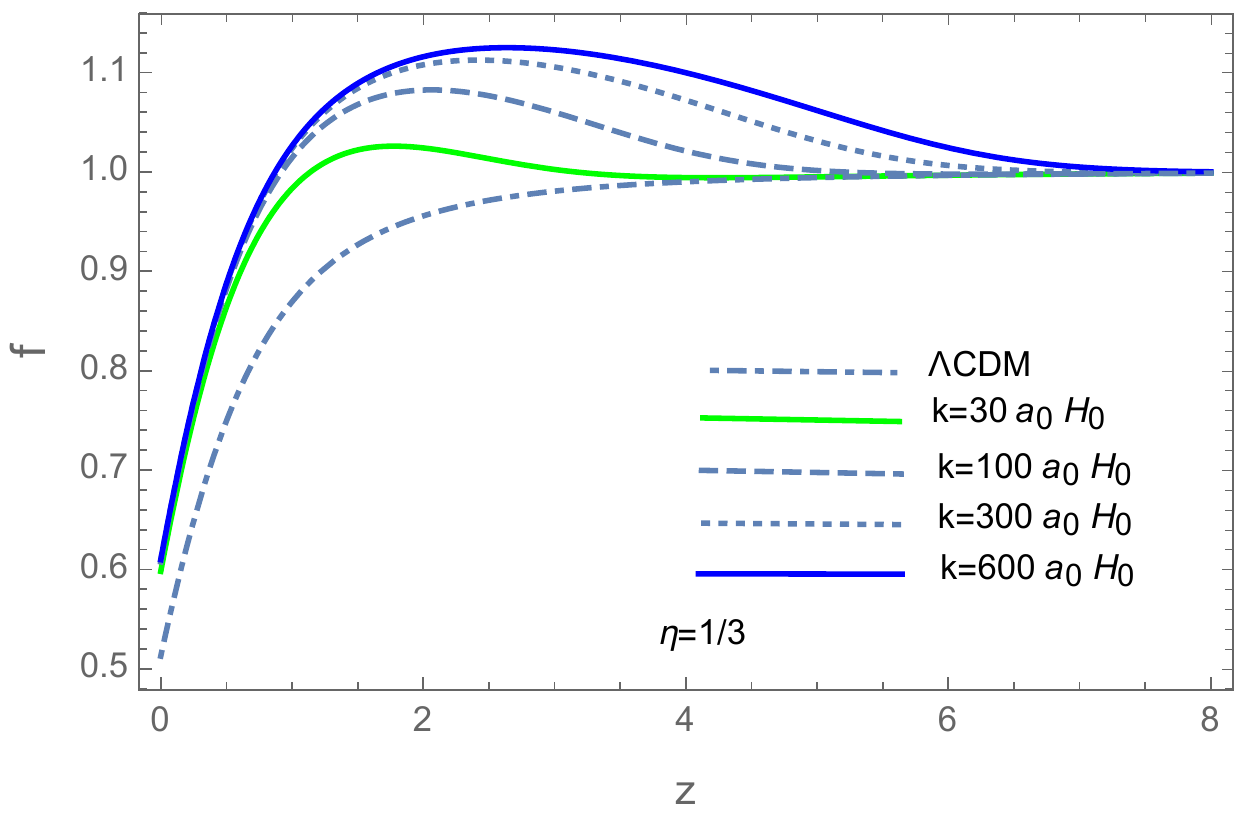}
\includegraphics[scale=0.59]{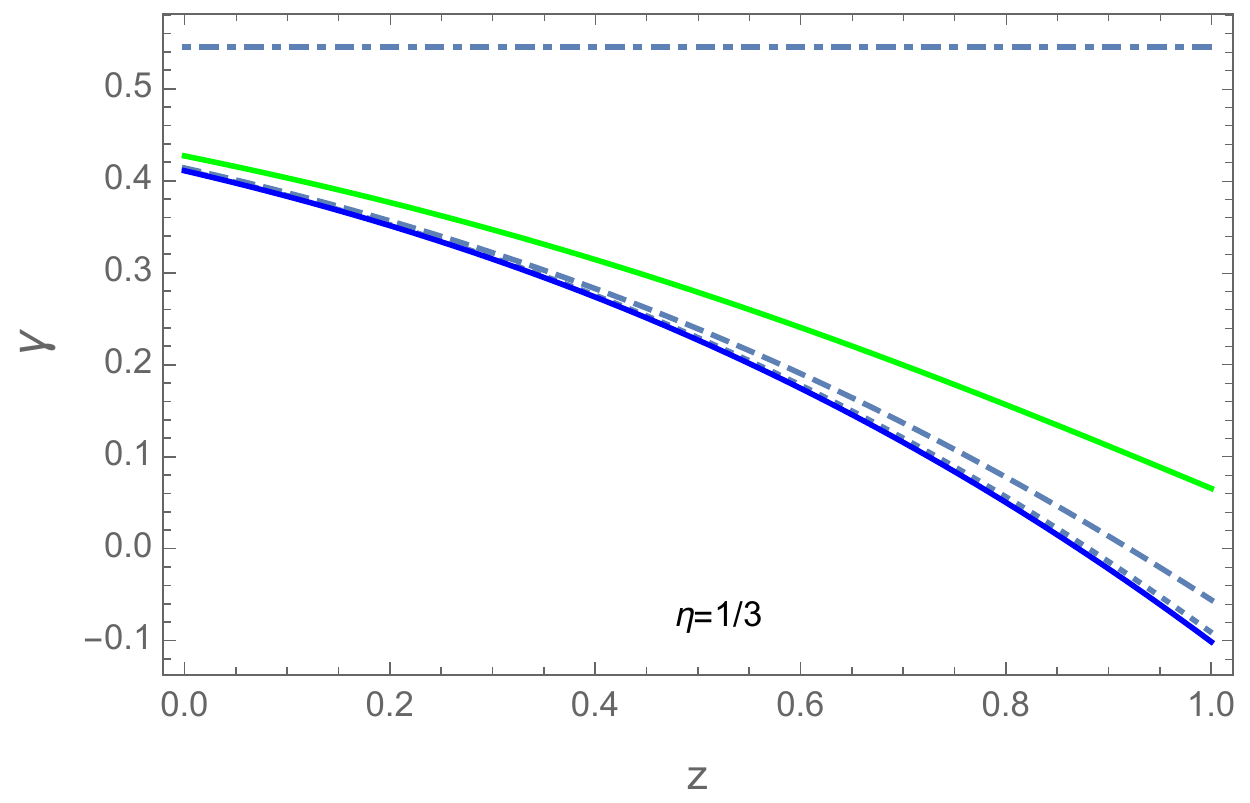}
\caption{The evolution of the growth rate $f$ and the growth index $\gamma$ for the model (\ref{model-l1}) for four different values of $k$. The upper figure corresponds to $\eta=2/5$, while the figure below corresponds to $\eta=1/3$.}
\label{fig5}
\end{center}
\end{figure}
As seen from Fig. 5, the transition to scalar-tensor regime begins quite early in the matter-dominated epoch, and the dispersion at low redshift is practically absent in all the curves, so that the asymptotic regime $k>>aM$ has already been reached at current epoch.
\noindent The bound $m\gtrsim 10^{-6}$ is satisfied for $\eta<3/4$. 
Note that the larger values of the deviation parameter $m$ compared to the simple exponential with $\eta=1$, allow the implementation of the chameleon mechanism for the model (\ref{model-l1}), although the extremely small Compton wavelength does not contradict the observations.\\

\noindent {\bf The $f\sigma_8$ Tension.}\\

\noindent  An important cosmological test for dark energy and modified gravity models that has been intensively analyzed  lately is the weighted growth rate, expressed as $f\sigma_8(a)$, in connection with the discordance found between
CMB and LSS observations.  $\sigma_8(a)$ is the matter `power spectrum normalization on scales of $8h^{-1}Mpc$, and the product $f\sigma_8(a)$ is independent of the bias factor between the observed galaxy spectrum and the underlying matter power spectrum \cite{percival}.
The values of $\sigma_8$ predicted by the $\Lambda$CDM model lead to an exceeding structure formation power compared to LSS observations. 
The weighted growth rate is expressed as
\be\label{sigma8}
f(a)\sigma_8(a)=\frac{\sigma_8}{\delta(1)}a\delta'(a)=-\frac{\sigma_8}{\sigma(0)}(1+z)\frac{d\delta (z)}{dz}
\ee
where $\sigma_8(a)=\sigma_8 \delta(a)/\delta(1)$ is the r.m.s. fluctuation of density perturbations on scale  $8h^{-1}Mpc$ and $\sigma_8$ is its current value.
In redshift variable the Eq. (\ref{matter-pert}) reads
\be\label{deltam}
(1+z)^2\delta''(z)+(1+z)\left[\frac{3}{2}\Omega_m(z)-1\right]\delta'(z)-\frac{3}{2}\frac{G_{eff}}{G}\Omega_m(z)\delta(z)=0
\ee
where, according to the results illustrated in Figs. 1 and 2, the $\Lambda$CDM background evolution can be assumed.
Numerical solution of this equation together with Eq. (\ref{sigma8}), with initial conditions in the deep matter era $\delta(z_i)\sim 1/z_i$ and $\delta'(z_i)\approx 0$ (with $z_i\sim 50-100$), gives the theoretical prediction of the models (\ref{model2}) and (\ref{moldel-l})  for $f\sigma_8(z)$. 
In Figs. 6 and 7 we plot the theoretical results for the weighted growth rate $f(z)\sigma_8(z)$ from the models (\ref{model2}) and (\ref{moldel-l})  respectively, contrasted with the data set from independent $f\sigma_8$ measurements from various surveys \cite{percival1, davis, hudson, turnbull, samushia, blake, blake1, ariel, chia1, howlett, feix, okumura, huterer1, pezzotta}. For the theoretical curves we have assumed $\Omega_{m0}=0.3$ and $\sigma_8=0.82$ consistent with Planck15/$\Lambda$CDM data.
 \begin{figure}
\begin{center}
\includegraphics[scale=0.9]{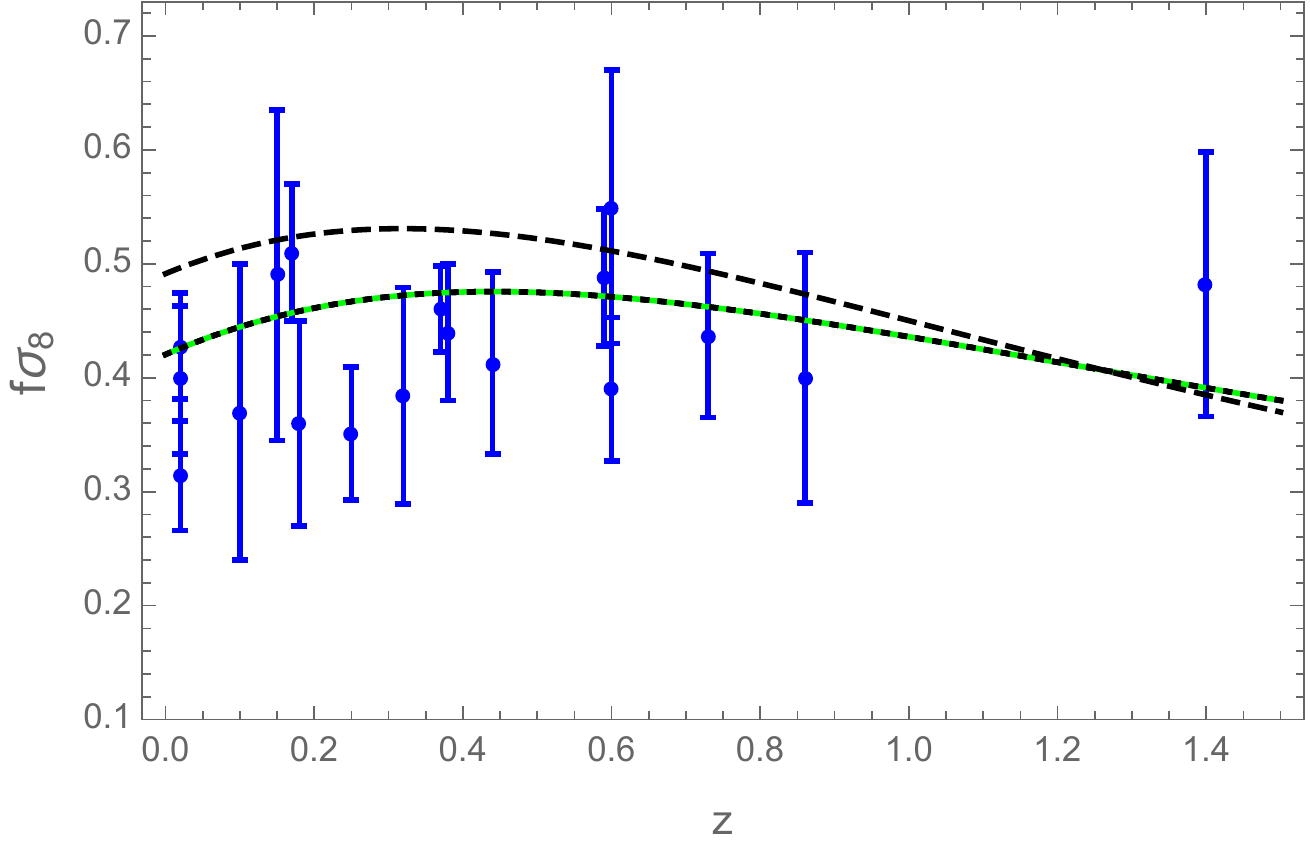}
\caption{The evolution of $f\sigma_8$ for the model (\ref{model2}). The different curves are calculated for the mode $k=300 a_0H_0$, taking $\Omega_{m0}=0.30$. The dashed line corresponds to $\eta=3$ and $\alpha=10$, and the $\Lambda$CDM corresponds to the green line. The remaining curves, that correspond to $\eta=4,5,6$ and respectively $\alpha=10^2,10^4, 10^6$, are indistinguishable form the $\Lambda$CDM model. Therefore for $\eta\ge 4$ the model aligns with the cosmological constant and does not relax the $f\sigma_8$ tension.}
\label{fig6}
\end{center}
\end{figure}

 \begin{figure}
\begin{center}
\includegraphics[scale=0.9]{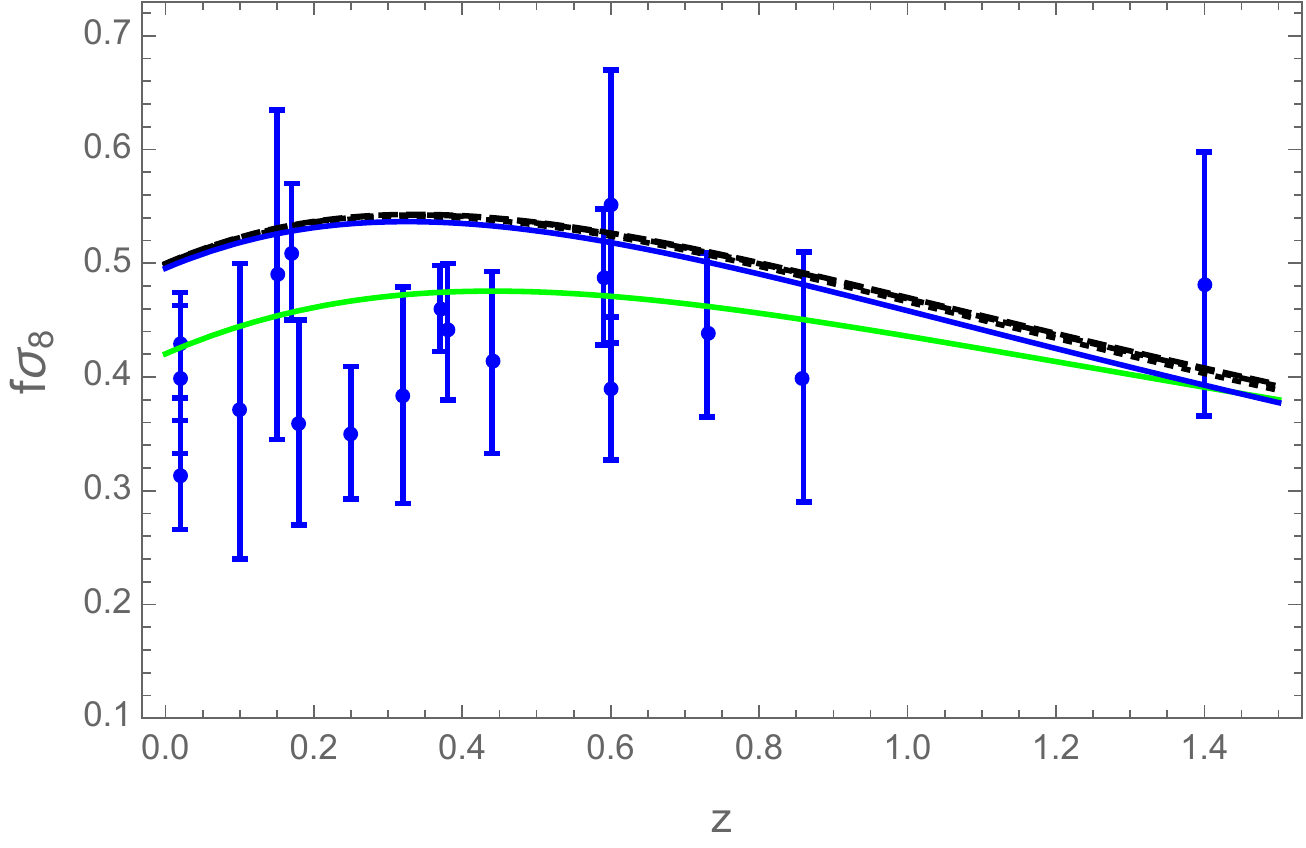}
\caption{The evolution of $f\sigma_8$ for the model (\ref{moldel-l}) for $\eta=2/7,1/3,2/5,1/2$, contrasted with the observed values of $f\sigma_8$ from the cited surveys. Note the difficulty of the models in adjusting to measurements of $f\sigma_8$ below 0.4 at low redshifts. The curves correspond to the mode $k=300 a_0H_0$, assuming $\Omega_{m0}=0.3$ and initial conditions for Eq. (\ref{deltam}) at $z_i=50$. The solid curve corresponds to $\eta=1/2$ and the green curve corresponds to $\Lambda$CDM.}
\label{fig6}
\end{center}
\end{figure}
\noindent  It is clear from Figs. 4 and 5 that the evolution of the growth index would allow us to distinguish the analyzed models from the $\Lambda$CDM model. However, concerning the $f\sigma_8$ evolution at low redshifts, the models show  exceeding structure formation power compared to LSS observations.\\
\noindent The above considered models behave asymptotically as $\Lambda$CDM as $R\to \infty$ and therefore, like $\Lambda$CDM, cannot give rise to inflation. However, if we add for instance the widely known $R^2$-term \cite{starobinsky} that is compatible with the latest Planck data \cite{planck18X}, then this term will dominate at curvatures typical of the inflation regime and with the adequate coupling will be irrelevant at late times. Then adding the term $\frac{R^2}{6M^2}$ where $M\approx 10^{-5}M_p\sim 10^{13}$ Gev, at large curvature one can neglect the dark energy component and the dominant Lagrangian becomes
\be\label{infl}
f(R)=R+\frac{R^2}{6M^2}
\ee
which gives the well known results for the scalar spectral index $n_s $ and the tensor-to-scalar ratio $r$:
\be\label{inflation1}
n_s-1\simeq -\frac{2}{N}, \;\; r\simeq \frac{12}{N^2}.
\ee
Then, by adding the above $R^2$-term to the each of the models  (\ref{model2}), (\ref{model-l1}), we can have a complete and consistent model unifying early time inflation with late time accelerated expansion. Starobinsky type inflation can also be accomplished if we multiply $R^2$ by the  factor $e^{-\left(\frac{\mu^2}{R}\right)^{\eta}}$, where $\eta<1$ and $R>>\mu^2$ during inflation \cite{granda3}.\\
\section{Discussion}
\noindent 
The main challenge of the modified gravity models in the explanation of late time accelerated expansion is the fulfillment of cosmological and the strictest local gravity constraints, while maintaining its own signatures that differentiate them from the standard $\Lambda$CDM model. In the present paper we propose models that comply with all these requirements and can lead to measurable signs of scalar-tensor regime from matter density perturbations. 
Two types of models were considered: models of the type $f(R)=R-\lambda\mu^2 e^{-g_1(R)}$, where the positive definite function $g_1(R)$ satisfies the asymptotic behavior $g_1(R\to \infty)\to 0$ and $g_1(R\to 0)\to \infty$, and models of the type $f(R)=R-\lambda\mu^2\left(1- e^{-g_2(R)}\right)$ where the positive definite function $g_2(R)$ behaves as $g_2(R\to \infty)\to \infty$ and $g_2(R\to 0)\to 0$.
The first limit leads to an effective cosmological constant while the second leads to disappearing cosmological constant in the flat space time limit.\\
Particularly we considered the models (\ref{model2}) and (\ref{model-l1}). In Figs. 1 and 2 we show the background evolution of the models, where for the assumed parameters, they present a mildly phantom behavior in the past, being more marked for the model (\ref{model-l1}), and as $\eta$ increases, the redshift at the phantom divide crossing decreases in both models. The parameter $\alpha$ in (\ref{model2}) can increase the steepness of $m$ at late times, favoring compliance with the bound $m(z\approx 0)\gtrsim 10^{-6}$ (see table 1), while at high redshift or large curvature the leading parameter is $\eta$. Since $\eta<1$ in the model (\ref{model-l1}), it can 
substantially attenuate the rapid zero trend at high redshift of the deviation parameter $m$, compared to the case $\eta = 1$, while respecting all local gravity and cosmological restrictions. This model gives a wide range of $m$-values that satisfy the bound $m(z\approx 0)\gtrsim 10^{-6}$. Additionally, given that $\eta<1$ in the model (\ref{model-l1}) allows to apply the chameleon mechanism that is absent in the case $\eta=1$ since the Compton wavelength becomes extremely short.\\
Testing the models under local gravity experiments it was found that $\eta>2.4$ for the model (\ref{model2}) (under the initial condition $f_{,R_0}=10^{-2}$), and $\eta>0.22$ for the model (\ref{model-l1}). Despite these bounds, the models may present appreciable deviations from the $\Lambda$CDM model.
Analyzing the evolution of matter density perturbations both models leave imprints that differentiate them from the $\Lambda$CDM model. 
For wave numbers relevant to the linear regime of the galaxy power spectrum, both models show values of the deviation parameter $m$ at current epoch (see tables I and II) that are larger than the lower limit established for the transition ($m\gtrsim 3\times 10^{-6}$), which could be between the scope of near future observations. There is also a transition from GR to ST regime in matter era, in the linear regime of perturbations, that is characterized by the redshift $z_k$. Some cases for the mode $k/(a_0H_0)=300$ are shown in tables I and  II, where for the model (\ref{model-l1}) the results show that as $\eta$ increases the transition is delayed.\\ 
The dispersion of $\gamma$ at low redshift for the model (\ref{model2}) (illustrated in Fig. 4)  is a clear signature of the scale dependence of $G_{eff}(k)$, which is a signal of modified gravity, but even if dispersion at current epoch is not present, as in the case of the model (\ref{model-l1}) (Fig. 5), the large departure of $\gamma(z=0)$ from $\gamma_0$ is clear signal of modified gravity. \\
The proposed models possess rich observational signatures relevant to SN Ia, galaxy clustering and CMB, and can show appreciable deviation from the $\Lambda$CDM model around the present epoch. However, this departure form $\Lambda$CDM does not relax the
$\sigma_8$-tension, since the models (\ref{model2}) and (\ref{model-l1}) give an exceeding structure formation power compared to $\Lambda$CDM. This is due to the fact that $G_{eff}$ increases from its Newtonian value $G$ to $4G/3$ through the evolution from matter dominated epoch to the present epoch of accelerated expansion. To relax the $\sigma_8$-tension in the frame of modified gravity it would require 
 physical interactions that lead to decreasing $G_{eff}$ from matter dominated to current epoch. 

\section*{Acknowledgments}
This work was supported by Fondo Nacional de Financiamiento para la Ciencia, la Tecnología y la Innovación Francisco José de Caldas (MINCIENCIAS - COLOMBIA) Grant No. 110685269447 RC-80740-465-2020, projects 69723 and 69553.


\begin{thebibliography}{99}

\bibitem{copeland} E. J. Copeland, M. Sami and S. Tsujikawa, Int. J. Mod. Phys. D \textbf{15}
1753-1936 (2006), arXiv:hep-th/0603057
\bibitem{sahnii} V. Sahni, 	Lect. Notes Phys. \textbf{653}, 141-180 (2004), arXiv:astro-ph/0403324v3
\bibitem{padmanabhan} T. Padmanabhan, Phys. Rept. \textbf{380}, 235 (2003), [hep-th/0212290].
\bibitem{sergeiod} K. Bamba, S. Capozziello, S. Nojiri, S. D. Odintsov, Astrophys.  and Space Sci. {\bf 342}, 155  (2012); arXiv:1205.3421 [gr-qc] 
\bibitem{ligo} B. P. Abbott et al. (Virgo, LIGO Scientific), Phys. Rev. Lett. {\bf 119}, 141101 (2017), 1709.09660.
\bibitem{horndeski} G. W. Horndeski, Int. J. Theor. Phys. {\bf 10}, 363 (1974).
\bibitem{nicolis} A. Nicolis, R. Rattazzi, E. Trincherini, Phys. Rev. D {\bf 79}, 064036 (2009); arXiv:0811.2197 [hep-th]
\bibitem{deffayet} C. Deffayet, G. Esposito-Farese, A. Vikman, Phys. Rev. D {\bf 79}, 084003 (2009), arXiv:0901.1314 [hep-th]
\bibitem{sodintsov1} S. Nojiri and S. D. Odintsov, Int. J. Geom. Meth. Mod. Phys. {\bf 4}, 115 (2007) [arXiv:hep-th/0601213].
\bibitem{sotiriou} T. P. Sotiriou, V. Faraoni, Rev. Mod. Phys. \textbf{82}, 451 (2010); arXiv:0805.1726 [gr-qc].
\bibitem{tsujikawa01} A. De Felice, S. Tsujikawa, Living Rev. Rel. {\bf 13}, 3  (2010); arXiv:1002.4928 [gr-qc]
\bibitem{tsujikawa0} S. Tsujikawa, Lect. Notes Phys. {\bf 800}, 99 (2010); arXiv:1101.0191 [gr-qc]
\bibitem{odinnojiri} S. Nojiri, S. D. Odintsov, Phys. Rept. {\bf 505} (2011) 59-144; arXiv:1011.0544 [gr-qc] (unified)
\bibitem{odinoiko4} S. Nojiri, S. D. Odintsov, V. K. Oikonomou, Phys. Rept. {\bf 692} (2017) 1-104; arXiv:1705.11098 [gr-qc]
\bibitem{capozziello} S. Capozziello, Int. J. Mod. Phys. D \textbf{11}, 483 (2002); gr-qc/0201033
\bibitem{capozziello1} S. Capozziello, S. Carloni, A. Troisi, Recent Res. Dev. Astron. Astrophys {\bf 1}, 625 (2003); astro-ph/0303041
\bibitem{sodintsov} S. Nojiri and S.D. Odintsov,Phys. Lett. B \textbf{576}, 5 (2003); hep-th/0307071
\bibitem{nojiri5} S. Nojiri and S. D. Odintsov, Phys. Rev. D {\bf 68}, 123512 (2003); arXiv:hep-th/0307288.
\bibitem{carroll} S. M. Carroll, V. Duvvuri, M. Trodden and M. S. Turner, Phys. Rev. D {\bf 70},
043528 (2004); arXiv:astro-ph/0306438
\bibitem{nojiriodin}   S. Nojiri and S. D. Odintsov, Gen. Rel. Grav. {\bf 36}, 1765 (2004), hep-th/0308176.
\bibitem{nojiri} M. C. B. Abdalla, S. Nojiri, and S. D. Odintsov, Class. Quant. Grav. \textbf{22}, L35 (2005); hep-th/0409177.
\bibitem{elizalde} G. Cognola, E. Elizalde, S. Nojiri, S. D. Odintsov, and S. Zerbini, JCAP {\bf 0502}, 010 (2005); hep-th/0501096.
\bibitem{troisi} S. Capozziello, V. F. Cardone, and A. Troisi, Phys. Rev. D \textbf{71}, 043503 (2005); astro-ph/0501426
\bibitem{allemandi} G. Allemandi, A. Borowiec, M. Francaviglia, and S. D. Odintsov, Phys. Rev. D {\bf 72}, 063505 (2005); grqc/0504057
\bibitem{koivisto} T. Koivisto and H. Kurki-Suonio, Class. Quant. Grav. \textbf{23}, 2355 (2006); astro-ph/0509422.
\bibitem{msami}  M. Sami, A. Toporensky, P. V. Tretjakov, and S. Tsujikawa, Phys. Lett. B{\bf 619}, 193 (2005), hep-th/0504154.
\bibitem{barrow1a} T. Clifton and J. D. Barrow, Phys. Rev. D{\bf 72}, 103005 (2005), gr-qc/0509059.
\bibitem{faraoni} V. Faraoni, Phys. Rev. D {\bf 72}, 124005 (2005); gr-qc/0511094.
\bibitem{brevik} I. Brevik, Int. J. Mod. Phys. D {\bf 15}, 767 (2006); gr-qc/0601100
\bibitem{koivisto1} T. Koivisto, Phys. Rev. D{\bf 73}, 083517 (2006), astro-ph/0602031.
\bibitem{sotiriou1} T. P. Sotiriou, Class. Quant. Grav. {\bf 23}, 5117 (2006), gr-qc/0604028.
\bibitem{nojiri1} S. Capozziello, S. Nojiri, S. D. Odintsov, and A. Troisi, Phys. Lett. B {\bf 639}, 135 (2006); astro-ph/0604431
\bibitem{dobado} A. de la Cruz-Dombriz and A. Dobado, Phys. Rev. D {\bf 74}, 087501 (2006); gr-qc/0607118.
\bibitem{sodintsov2} S. Nojiri and S. D. Odintsov, Phys. Rev. D {\bf74}, 086005 (2006); hep-th/0608008
\bibitem{anthoni} A. W. Brookfield, C. Van de Bruck, and L. M. H. Hall, Phys. Rev. D {\bf 74}, 064028 (2006); hep-th/0608015.
\bibitem{nojiri2} S. Nojiri and S. D. Odintsov (2006), J. Phys. A {\bf 40}, 6725 (2007); hep-th/0610164
\bibitem{faraoni1} V. Faraoni, Phys. Rev. D{\bf 74}, 104017 (2006), astro-ph/0610734.
\bibitem{song}  Y.-S. Song, W. Hu, and I. Sawicki, Phys. Rev. D{\bf 75}, 044004 (2007), astro-ph/0610532.
\bibitem{bean} R. Bean, D. Bernat, L. Pogosian, A. Silvestri, M. Trodden, Phys. Rev. D{\bf 75}, 064020 (2007), astro-ph/0611321.
\bibitem{olmo} G. J. Olmo, Phys. Rev. D {\bf 75}, 023511 (2007); gr-qc/0612047.
\bibitem{amendola1} L. Amendola, R. Gannouji, D. Polarski and S. Tsujikawa, Phys. Rev. D\textbf{75}, 083504 (2007); [arXiv:gr-qc/0612180].
\bibitem{barrow1} B. Li and J. D. Barrow, Phys. Rev. D {\bf75}, 084010 (2007); gr-qc/0701111.
\bibitem{fay}  S. Fay, S. Nesseris, and L. Perivolaropoulos, Phys. Rev. D{\bf 76}, 063504 (2007), gr-qc/0703006.
\bibitem{faraoni2} V. Faraoni, Phys. Rev. D{\bf 75}, 067302 (2007), gr-qc/0703044.
\bibitem{hu} W. Hu and I. Sawicki, Phys. Rev. D \textbf{76}, 064004 (2007); [arXiv:astro-ph/0705.1158].
\bibitem{nojiri6} S. Nojiri and S. D. Odintsov, Phys. Lett. B {\bf 657}, 238 (2007); arXiv: 0707.1941 [hep-th].       
\bibitem{tsujikawa1} S. Tsujikawa, Phys. Rev. D {\bf 77}, 023507 (2008); arXiv:0709.1391 [astro-ph]
\bibitem{nojiri7} S. Nojiri and S. D. Odintsov, Phys. Rev. D {\bf 77}, 026007 (2008); arXiv:0710.1738 [hep-th].    
\bibitem{sergeid2} G. Cognola, E. Elizalde, S. Nojiri, S.D. Odintsov, L. Sebastiani, S. Zerbini, Phys. Rev. D {\bf 77}, 046009 (2008); arXiv:0712.4017 [hep-th]. 

\bibitem{bamba1} K. Bamba, S. Nojiri, S.D. Odintsov, JCAP {\bf 0810}, 045 (2008); [arXiv:0807.2575 [hep-th].
\bibitem{elizalde1} E. Elizalde, S. Nojiri, S. D. Odintsov, L. Sebastiani, S. Zerbini, Phys. Rev. D{\bf 83}, 086006 (2011); arXiv:1012.2280 
\bibitem{sergeioiko} S. D. Odintsov, V. K. Oikonomou, Nucl. Phys. B{\bf 293}, 608 (2017); arXiv:1708.08346.
\bibitem{sergeisaez} S. D. Odintsov, D. Saez-Chillon, G. S. Sharov, Eur. Phys. J. C{\bf 77}, 862 (2017); arXiv:1709.06800
\bibitem{sergeisaez1} S. D. Odintsov, D. Saez-Chillon, G. S. Sharov, Phys. Rev. D{\bf 99}, 024003 (2019); arXiv:1807.02163.  
\bibitem{starobinsky} A. A. Starobinsky, Phys. Lett. B \textbf{91}, 99 (1980).
\bibitem{dolgov} A. D. Dolgov and M. Kawasaki, Phys. Lett. B {\bf 573}, 1 (2003); [arXiv:astro-ph/0307285].
\bibitem{lucamendola} L. Amendola, D. Polarski and S. Tsujikawa, Phys. Rev. Lett. {\bf98}, 131302 (2007) [arXiv:astro-ph/0603703].   
\bibitem{olmo1} G. J. Olmo, Phys. Rev. D {\bf 72}, 083505 (2005); arXiv:gr-qc/0505135
\bibitem{troisi1} S. Capozziello, V. F. Cardone, S. Carloni and A. Troisi, Int. J. Mod. Phys. D {\bf 12}, 1969 (2003), arXiv:astro-ph/0307018.
\bibitem{mota} M. Amarzguioui, O. Elgaroy, D. F. Mota and T. Multamaki, Astron. Astrophys. {\bf 454}, 707 (2006), arXiv:astro-ph/0510519.
\bibitem{dev} A. Dev, D. Jain, S. Jhingan, S. Nojiri, M. Sami and I. Thongkool, Phys. Rev. D {\bf 78}, 083515 (2008), arXiv:0807.3445.
\bibitem{schmidt} F. Schmidt, A. Vikhlinin and W. Hu, Phys. Rev. D {\bf 80}, 083505 (2009), [arXiv:0908.2457].
\bibitem{lombriser} L. Lombriser, A. Slosar, U. Seljak and W. Hu, Phys. Rev. D {\bf 85}, 124038 (2012), arXiv:1003.3009.
\bibitem{nesseris} S. Basilakos, S. Nesseris and L. Perivolaropoulos, Phys. Rev. D {\bf 87}, 123529 (2013), arXiv:1302.6051.
\bibitem{nunes} R. C. Nunes, S. Pan, E. N. Saridakis, E. M. C. Abreu, JCAP {\bf 1701},  005 (2017), arXiv:1610.07518 [astro-ph.CO]
\bibitem{chiba} T. Chiba, T. L. Smith and A. L. Erickcek, Phys. Rev. D {\bf 75}, 124014 (2007), arXiv:astro-ph/0611867.
\bibitem{lamendola5} L. Amendola and S. Tsujikawa, Phys. Lett. B {\bf 660}, 125 (2008), arXiv:0705.0396.
\bibitem{tsujikawa} S. Capozziello and S. Tsujikawa, Phys. Rev. D {\bf77}, 107501 (2008); arXiv:0712.2268.
\bibitem{brax} P. Brax, C. van de Bruck, A. Davis, and D. J. Shaw. Phys. Rev. D {\bf 78}, 104021 (2008); arXiv:0806.3415.
\bibitem{astarobinsky} A. A. Starobinsky, JETP Lett. {\bf 86}, 157 (2007); arXiv:0706.2041 [astro-ph]
\bibitem{appleby1} S. A. Appleby and R. A. Battye, Phys. Lett. {\bf B 654}, 7 (2007); arXiv:0705.3199 [astro-ph]
\bibitem{granda1} L. N. Granda, Eur. Phys. J. C. {\bf 80}, 538 (2020); arXiv:2003.09006 [gr-qc]
\bibitem{granda3} L. N. Granda, Symmetry {\bf 12}, 794 (2020)
\bibitem{amanda1} J. Khoury, A. Weltman, Phys. Rev. Lett. {\bf 93}, 171104 (2004); arXiv:astro-ph/0309300
\bibitem{amanda2} J. Khoury, A. Weltman, Phys. Rev. D {\bf 69}, 044026 (2004); arXiv:astro-ph/0309411
\bibitem{linder1} E. V. Linder, Phys. Rev. D {\bf 80}, 123528 (2009); arXiv:0905.2962 [astro-ph.CO]
\bibitem{tegmark} T. Faulkner, M. Tegmark, E. F. Bunn and Y. Mao, Phys. Rev. D {\bf 76}, 063505 (2007), arXiv:astro-ph/0612569

\bibitem{cmwill} C. M. Will, Living Rev. Relativity {\bf 9}, 3 (2006), arXiv:gr-qc/0510072.
\bibitem{tsujikawa10} S. Tsujikawa, Phys.  Rev.D {\bf 76}, 023514 (2007); arXiv:0705.1032 [astro-ph]
\bibitem{tsujikawa11} S.Tsujikawa, K. Uddin, R. Tavakol, Phys. Rev. D {\bf 77}, 043007 (2008); arXiv:0712.0082 [astro-ph]
\bibitem{craig} C. J. Copi, A. N. Davis, L. M. Krauss, Phys. Rev. Lett.  {\bf 92}, 171301 (2004); arXiv:astro-ph/0311334
\bibitem{turner} S. Burles, K. M. Nollett, M. S. Turner, Phys. Rev.  D {\bf 63}, 063512 (2001); arXiv:astro-ph/0008495
\bibitem{serpico} F. Iocco, G. Mangano, G. Miele, O. Pisanti, P.D. Serpico, Phys. Rept. {\bf 472}, 1-76 (2009); 0809.0631 [astro-ph]
\bibitem{scarpetta} G. Lambiase, G. Scarpetta, Phys. Rev.  D {\bf 74}, 087504 (2006); arXiv:astro-ph/0610367
\bibitem{neseris} S. Nesseris, A. Mazumdar, Phys. Rev. D {\bf 79},104006 (2009).

\bibitem{tegmark1} Tegmark, M. et al. (SDSS Collaboration), Cosmological parameters from SDSS and
WMAP, Phys. Rev. D, 69, 103501, (2004).
\bibitem{tegmark2} Tegmark, M. et al. (SDSS Collaboration), Cosmological constraints from the SDSS luminous
red galaxies, Phys. Rev. D, 74, 123507, (2006).
\bibitem{linder} E. V. Linder, Phys. Rev. D {\bf 72}, 043529 (2005); arXiv:astro-ph/0507263
\bibitem{percival} W. J. Percival et. al, Astrophys. J. {\bf 657}, 645 (2007).
\bibitem{huterer} D. Huterer and E. V. Linder, Phys. Rev. D {\bf 75}, 023519 (2007); arXiv:astro-ph/0608681
\bibitem{tsujikawa3} S.Tsujikawa, R. Gannouji, B. Moraes, D. Polarski, Phys. Rev. D {\bf 80}, 084044 (2009); arXiv:0908.2669 [astro-ph.CO]
\bibitem{peebles} P. J. E. Peebles, Astrophys. J.284, 439 (1984).
\bibitem{wangstein}   L. Wang, P. J. Steinhardt, Astrophys. J. {\bf 508}, 483  (1998); arXiv:astro-ph/9804015
\bibitem{percival1} Y-S. Song, W. J. Percival, JCAP {\bf 0910}, 004 (2009), arXiv:0807.0810 [astro-
ph].
\bibitem{davis} M. Davis, A. Nusser, K. Masters, C. Springob, J. P. Huchra, G. Lemson, Mon. Not. Roy. Astron. Soc. {\bf 413}, 2906
(2011), arXiv:1011.3114 [astro-ph.CO].
\bibitem{hudson} M. J. Hudson, Stephen J. Turnbull, The Astrophysical Journal Letters {\bf 751}, L30 (2012); 1203.4814 [astro-ph.CO].
\bibitem{turnbull} S. J. Turnbull, M. J. Hudson, H. A. Feldman,
M. Hicken, R. P. Kirshner, R. Watkins,  Mon. Not. Roy. Astron. Soc. 420, 447 (2012), arXiv:1111.0631 [astro-ph.CO].
\bibitem{samushia} L. Samushia, W. J. Percival, and A. Raccanelli, Mon. Not. Roy. Astron. Soc.
{\bf 420}, 2102 (2012), arXiv:1102.1014 [astro-ph.CO].
\bibitem{blake} C. Blake et. al., Mon. Not. Roy. Astron. Soc. {\bf 425}, 405 (2012), arXiv:1204.3674 [astro-ph.CO].
\bibitem{blake1} C. Blake et al.,  Mon. Not. Roy. Astron. Soc. {\bf 436}, 3089 (2013), arXiv:1309.5556 [astro-ph.CO].
\bibitem{ariel} A. G. Sanchez et al., Mon. Not. Roy. Astron. Soc. {bf 440}, 2692 (2014), arXiv:1312.4854 [astro-ph.CO].
\bibitem{chia1} Chia-Hsun Chuang et al., Mon. Not. Roy. Astron. Soc. {\bf 461}, 3781 (2016), arXiv:1312.4889 [astro-ph.CO].
\bibitem{howlett} C. Howlett, A. Ross, L. Samushia, W. Percival, Marc Manera, Mon. Not. Roy. Astron.
Soc. {\bf 449}, 848 (2015), arXiv:1409.3238 [astro-ph. CO].
\bibitem{feix} M. Feix, A. Nusser, E. Branchini, Phys. Rev. Lett. {\bf 115}, 011301 (2015), arXiv:1503.05945 [astro-ph.CO].
\bibitem{okumura} T. Okumura et al., Publ. Astron. Soc. Jap. 68, 24 (2016), arXiv:1511.08083
[astro-ph.CO].
\bibitem{huterer1} D. Huterer, D. Shafer, D. Scolnic, F. Schmidt, JCAP {\bf 1705}, 015 (2017), arXiv:1611.09862 [astro-ph.CO].
\bibitem{pezzotta} A. Pezzotta et al., Astron. Astro-phys. 604, A33 (2017), arXiv:1612.05645 [astro-ph.CO].
\bibitem{planck18X} Y. Akrami et al., Planck Collaboration (Planck 2018 results. X. Constraints on inflation), arXiv:1807.06211 [astro-ph.CO]


\end{thebibliography}
\end{document}